\documentclass[fleqn,10pt]{wlscirep}
\usepackage[utf8]{inputenc}
\usepackage[T1]{fontenc}
\usepackage{multirow}
\usepackage{multicol}
\usepackage{stmaryrd}
\usepackage{amsmath,amsfonts}
\usepackage{algorithmic}
\usepackage{algorithm}
\usepackage{makecell}
\usepackage{listings}
\usepackage{xcolor}
\usepackage{subfig}
\usepackage{subcaption}
\usepackage{graphicx}
\usepackage{bm}
\usepackage{comment}

\newcommand{\workname}{Lancelot}

\title{\workname: Towards Efficient and Privacy-Preserving Byzantine-Robust Federated Learning within Fully Homomorphic Encryption}

\author[1]{Siyang Jiang}
\author[2]{Hao Yang}
\author[3]{Qipeng Xie}
\author[4,*]{Chuan Ma}
\author[5]{Sen Wang}
\author[1,*]{Guoliang Xing}

\affil[1]{ Department of Information Engineering, The Chinese University of Hong Kong, Hong Kong SRA, China}
\affil[2]{College of Computer Science and Technology, Nanjing University of Aeronautics and Astronautics, Nanjing, China}
\affil[3]{IoT Thrust, Information Hub, Hong Kong University of Science and Technology (Guangzhou), China}
\affil[4]{School of Computer Science, Chongqing University, Chongqing, China}
\affil[5]{2012 Lab, Huawei Technologies, Shenzhen, China}
\affil[*]{Corresponding author: chuan.ma@njust.edu.cn, glxing@ie.cuhk.edu.hk}

\newcommand{\pk}{\mathbf{pk}}
\newcommand{\sk}{\mathbf{sk}}
\newcommand{\ct}{\mathbf{ct}}
\newcommand{\evk}{\mathbf{evk}}

\keywords{Byzantine-Robust Federated Learning, Fully Homomorphic Encryption, Privacy Preserving}

\begin{abstract}

In sectors such as finance and healthcare, where data governance is subject to rigorous regulatory requirements, the exchange and utilization of data are particularly challenging. Federated Learning (FL) has risen as a pioneering distributed machine learning paradigm that enables collaborative model training across multiple institutions while maintaining data decentralization. This approach inherently heightens data privacy by sharing only model weights, rather than raw data. Despite its advantages, FL is vulnerable to adversarial threats, particularly poisoning attacks during model aggregation, a process typically managed by a central server. To counteract the vulnerabilities of traditional FL frameworks, Byzantine-robust federated learning (BRFL) systems have been introduced, which rely on robust aggregation rules to mitigate the impact of malicious attacks. However, in these systems, neural network models still possess the capacity to inadvertently memorize and potentially expose individual training instances. This presents a significant privacy risk, as attackers could reconstruct private data by leveraging the information contained in the model itself. Existing solutions fall short of providing a viable, privacy-preserving BRFL system that is both completely secure against information leakage and computationally efficient. To address these concerns, we propose \workname, an innovative and computationally efficient BRFL framework that employs fully homomorphic encryption (FHE) to safeguard against malicious client activities while preserving data privacy. \workname~features a novel interactive sorting mechanism called \textit{masked-based encrypted sorting}. This method successfully circumvents the multiplication depth limitations of ciphertext, ensuring zero information leakage. Furthermore, we incorporate cryptographic enhancements, such as Lazy Relinearization and Dynamic Hoisting, alongside GPU hardware acceleration, to achieve a level of computational efficiency that makes \workname~a viable option for practical implementation. Our extensive testing, which includes medical imaging diagnostics and widely-used public image datasets, demonstrates that \workname~significantly outperforms existing methods, offering more than a twenty-fold increase in processing speed, all while maintaining data privacy. The \workname~framework thus stands as a potent solution to the pressing issue of privacy in secure, multi-centric scientific collaborations, paving the way for safer and more efficient federated learning applications.

\end{abstract}
\begin{document}

\flushbottom
\maketitle

\thispagestyle{empty}

\section*{Introduction}
Machine Learning (ML) techniques have achieved notable success, contingent upon a crucial prerequisite for fully unlocking their potential, enabling the seamless interconnection of vast datasets and ensuring their broad accessibility to researchers. Nevertheless, certain domains, particularly healthcare and finance, grapple with constraints on data sharing owing to regulatory standards and privacy considerations. Consequently, institutions in these sectors are unable to amalgamate and disseminate their data, impeding advancements in research and model development. The potential to share data among institutions while upholding data privacy could lead to the creation of more robust and accurate models. However, due to data sitting across numerous institutions, the task of centralizing and aggregating this information poses a formidable challenge, entangled in legal and regulatory hurdles. Notably, the General Data Protection Regulation (GDPR)~\cite{European} imposes stringent limitations on the transference of personal data, including pseudonymized data, across different jurisdictions.

Federated Learning (FL) is emerging as a novel paradigm to resolve data governance without sharing private data. For instance, FL enables various healthcare providers to collaboratively conduct statistical analyses and develop machine learning models without exchanging the underlying datasets~\cite{froelicher2021truly,kalra2023decentralized,nguyen2022federated,li2021federated}. Only aggregated results or model updates are transferred, allowing each healthcare provider to define its data governance and maintain control over access to its patient-level data. In this way, each healthcare provider can define its data governance and retain control over access to its patient-level data. 
Therefore, FL provides opportunities to utilize large and diverse volumes of data distributed across multiple institutions. These opportunities can aid in developing and validating artificial intelligence algorithms, resulting in more accurate, unbiased, and generalizable clinical recommendations and expediting novel discoveries. Such advancements are particularly crucial in rare diseases or medical conditions, where the number of affected patients within a single institution is often insufficient to identify meaningful statistical patterns with adequate statistical power.

Due to its distributed pattern, FL is susceptible to adversarial manipulations stemming from malicious clients. These could either be fake clients introduced by an attacker or genuine clients that an attacker has compromised. For instance, malicious clients have the potential to corrupt the global model by contaminating their local training data, a tactic commonly referred to as data poisoning attacks~\cite{Biggio2012PoisoningAA,Nelson2008ExploitingML}. Alternatively, they could tamper with their local model updates forwarded to the server, a strategy known as local model poisoning attacks~\cite{Fang2019LocalMP,Bhagoji2018AnalyzingFL,Bagdasaryan2018HowTB,Xie2020DBADB}. The corruption of the global model can lead to erroneous predictions encompassing a significant number of test examples indiscriminately, a circumstance that has severe ramifications, especially within medical applications~\cite{alkhunaizi2022suppressing}. Conversely, the compromised model may deliver predictions of attacker-specified target labels for test examples chosen by the attacker. In contrast, the predicted labels for other non-target test examples remain unaltered. Byzantine-Robust FL (BRFL) methods~\cite{Blanchard2017MachineLW,Chen2017DistributedSM,Mhamdi2018TheHV,Yang2019ByzantineResilientSG,Yin2018ByzantineRobustDL} aim to address malicious clients. The objective in this context is to learn an accurate global model even when a bounded number of clients engage in malicious activities. The core concept is to employ Byzantine-robust aggregation rules, primarily comparing the clients' local model updates and eliminating statistical outliers, e.g., comparing the distance among all weights, before utilizing them for global model updates. For example, BRFL has been deployed in thorax diseases with X-ray and pigmented skin lesions diagnosed~\cite{alkhunaizi2022suppressing} to avoid malicious client attacks and enjoy the benefits of knowing all participants without sharing their raw data.

Despite receiving commendations for its enhanced privacy protections, primarily attributed to the retention of raw data on the client's device, BRFL, can not guarantee the level of privacy that is essential for institutions operating within regulatory frameworks. BRFL involves each client disbursing gradient updates to the central server.  
This procedure introduces potential vulnerabilities, as deep neural networks harbor the capability to memorize individual training instances, potentially culminating in a comprehensive breach of the client's privacy. For example, DLG~\cite{zhu2019deep} and IDLG~\cite{zhao2020idlg} showed the possibility of stealing private training data from the shared gradients of other clients. To provide comprehensive privacy protection, more advanced solutions for FL have been proposed, such as Differential Privacy (DP)~\cite{wei2020federated}, Multiparty Computation (MPC)~\cite{bonawitz2017practical}, and Fully Homomorphic Encryption (FHE)~\cite{cheon2017homomorphic}. These techniques frequently achieve heightened privacy protection at the expense of accuracy or computational efficiency, constraining their applicability. 
Existing DP techniques for FL, which obstruct privacy leakage from intermediate data by introducing noise before sharing, often necessitate prohibitive amounts of noise, resulting in inaccurate models. Moreover, there is a lack of consensus on how to set the privacy parameters for DP to provide an acceptable mitigation of inference risks in practical scenarios. Conversely, MPC and FHE are cryptographic frameworks for securely executing computations over private datasets without intermediate leakage. 
However, MPC has significant drawbacks, as it incurs considerable network communication overhead and faces substantial challenges when scaling to accommodate a large number of data providers. Compared to DP and MPC, FHE-based BRFL systems offer a promising direction. They avoid inaccurate results and reduce multi-round communications. A vital feature of these systems is that all computations are encrypted, eliminating the need for third-party involvement. However, it should be noted that FHE-based BRFL systems generally incur a significant computational overhead. For example, during a single iteration, the computational delay in the context of semi-homomorphic encryption, which only supports the addition operation, is extended hundreds of times when processing ciphertext compared to plaintext~\cite{zhang2020batchcrypt}. This significant reduction in computational efficiency is even more evident in the case of FHE.


There are mainly two critical challenges to building an FHE-based BRFL system. The first challenge lies in the sorting operation in ciphertext due to the limitation of multiplication depth, which limits the deployment of BRFL aggregation rules. For example, some typical BRFL aggregation rules need to sort the distance according to the model weights, leading to multiple ciphertext comparisons during training. The second challenge is the computation efficiency of ciphertext, e.g., Add and Multiplication operations, resulting in prolonged training duration. In particular, the training time is generally increased thousands of times after adopting FHE. In fact, several open-source software platforms have recently been developed to provide users with streamlined access to FL algorithms, such as FedScale~\cite{oort-osdi21}, Flower~\cite{beutel2020flower}, NVFLARE~\cite{Roth2022NVIDIAFF} and FedML~\cite{he2020fedml}. It allows the user to federate any workload, any ML framework, and any programming language. However, none of the existing platforms address the problem of indirect privacy leakages in BRFL. It remains unclear whether these existing solutions can substantially simplify FHE-based BRFL systems compared to more conventional workflows that centralize the data if the local model updates could still be considered personal identifying data.

Therefore, in this paper, we present \workname, a computation-efficient framework for privacy-preserving byzantine-robust federated learning using fully homomorphic encryption.  In particular, we propose a novel interactive federated learning paradigm for decentralized collaboration between institutions, enabling high-performance and robust models to train without sacrificing data privacy and less computation efficiency.  \workname~incorporates algorithmic enhancements and hardware acceleration, encompassing optimizations of pair-wise ciphertext multiplication strategies, polynomial matrix multiplication, and complex operation additions. These techniques play a crucial role in the efficient design of Lancelot. Our contributions are as follows: (1) we propose a computation-efficient framework for  FL in multi-institutional collaborations that is adapted to heterogeneous data sources; (2) we incorporate FHE into BRFL to achieve rigorous privacy guarantees; (3) we analyze and reduce the computation overhead required to collaborate.

\section*{Results}

\subsection*{System Model}
As shown in Fig.~\ref{fig:Lacelot}, in this paper, there are three entities in \workname, including  \textit{Clients}, \textit{Server}, and  \textit{Key Generation Center}. In particular, \textit{Key Generation Center} is a trusted institution that generates public/private key pairs and distributes public keys to \textit{Clients}. Next, training data is stored in these \textit{Clients}, which only have public keys provided by the key generation center. The goal of \textit{Clients} is to benefit from a well-trained global model. 
Lastly, \textit{Server} receives the encrypted model from \textit{Clients} and performs the calculations in the ciphertext. 


\subsection*{Overview of \workname}
\begin{figure*}[tb]\small
  \centering{
  \includegraphics[width=\linewidth]{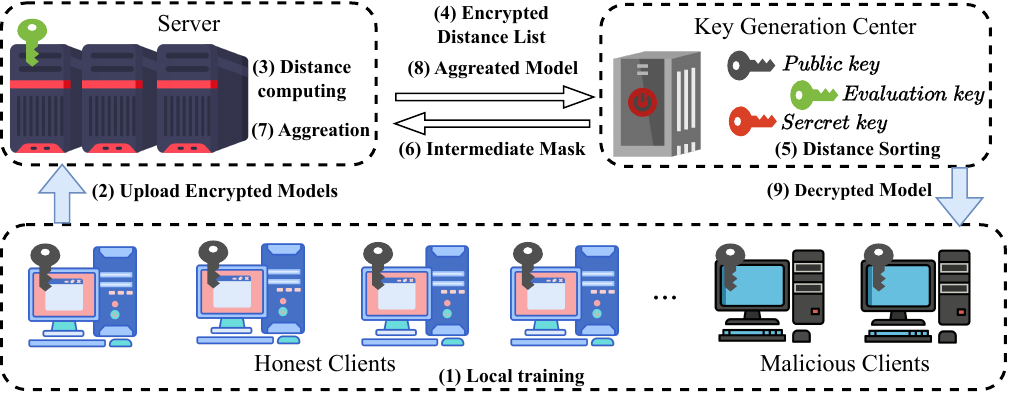}
  \caption{A workflow of \workname. At the beginning of \workname, the \textit{Key Generation Center} generates the cryptographic keys: the secret key $\sk$ is used for decryption of ciphertexts, and the public key $\pk$ is used for data encryption; the evaluation keys $\evk$ are used for homomorphic computations (e.g., ciphertext-ciphertext multiplications or ciphertext rotations). The $\pk$ is transmitted securely to the \textit{Clients}, and the $\evk$ is transmitted securely to the \textit{Server}. The \textit{Key Generation Center} needs to generate the keys, process the robust aggregation rules, and decrypt the aggregated model.  The \textit{Clients} encrypted models from plaintext and then transmitted to the \textit{Server}.  The models in \textit{Server} is processed in encrypted form, so it can only access the $\evk$ for homomorphic computation. The intermediate mask selects the \textit{Clients} according to the byzantine-robust aggregation rules in the cipher space.
  }
  \label{fig:Lacelot}
  }
\end{figure*}
As shown in Fig.~\ref{fig:Lacelot}, we provide an overview of the \workname~workflow.  We mainly have three stages of \workname.
\begin{itemize}
    \item \textbf{Stage 1:} Initially, clients encrypt their extensively trained models and transmit them to the server (Steps \textbf{(1)} and \textbf{(2)}). 
    \item \textbf{Stage 2:}  The server computes the distance between the models of different clients according to different aggregation rules, e.g., comparing the distance among all weights from clients (Step \textbf{(3)}).  It's important to note that this computation occurs in the ciphertext. Upon acquiring the list of distances, the server sends this information to the key generation center, where the encrypted distance list is decrypted and sorted (Steps \textbf{(4)} and \textbf{(5)}). Once the list is sorted, it is encoded into a distance mask and returned to the server for aggregation (Steps \textbf{(6)} and \textbf{(7)}). The index information of selected clients is encoded into the distance mask, enabling the server to perform the aggregation rules in an encrypted manner.
    \item \textbf{Stage 3:} Following the aggregation rules, the server combines the chosen models and forwards the aggregated model to the key generation center (Step \textbf{(8)}). In the final step (Step \textbf{(9)}), the key generation center employs the secret key to decrypt the encrypted model and distribute it to all clients. 
\end{itemize}

 Note that the model is decrypted and distributed to all clients during the final step without revealing individual client information. This ensures the confidentiality of each client's data throughout the process. We propose two acceleration strategies to mitigate the significant computational delay, e.g., cryptographic optimizations and hardware acceleration, which are detailed in the following  section of Methods. We also provide a pseudo-code of \workname~in the Appendix.


\subsection*{Baseline and Implementation details}
In order to evaluate the effectiveness of \workname,  we conduct experiments on a public image classification benchmark with four datasets. In addition, to verify its practical significance, we also conduct experiments on a medical image diagnosing benchmark with 12 datasets. We select three BRFL algorithms for comparison, e.g., Krum~\cite{Blanchard2017MachineLW}, Multi-Krum~\cite{Blanchard2017MachineLW}, Median~\cite{yin2018byzantine}. Krum aims to choose a single model from the pool
of $n$ local models, specifically favoring those that demonstrate high degrees of congruity with others in the set.  Multi-Krum extends the Krum algorithm to select multiple local model updates rather than one.  In Median~\cite{yin2018byzantine}, the server also sorts the values of each parameter and considers each parameter's median value in all local model updates. Note that we modify the conventional median method by choosing the median clients according to the distance. The detailed information of the above three methods can be referred to in the Appendix. 
In addition, we select vanilla BRFL and OpenFHE for implementations of Krum, Multi-Krum, and Median. In particular,  the vanilla BRFL implements these three algorithms in plaintext, and {OpenFHE~\cite{al2022openfhe}} is a state-of-the-art (SOTA) open-source FHE library that includes efficient implementations of typical FHE schemes. All experiments are conducted in LeNet-5~\cite{lecun1998gradient}, ResNet-18~\cite{he2016deep}, ResNet-34~\cite{he2016deep}, and ResNet-50~\cite{he2016deep} models. To ensure the accuracy of \workname, we also present the results between and FedAvg under the targeted attack~\cite{bhagoji2019analyzing} and untargeted attack, i.e., MAPF~\cite{cao2022mpaf}. 

We design and implement the Lancelot using NVIDIA GeForce RTX 4090 GPU and 4 * A100 on Ubuntu 20.04.6 LTS. 
Specifically, we allocate one process to manage server tasks, including compromised client selection and model aggregation. For each client, we establish a separate process responsible for local model updates and managing the data migration between CPUs and GPUs. Inter-process communication, executed in Python3, facilitates the interaction between the server and nodes. 
The deep learning components of our system are implemented using PyTorch. In all experiments, we use SGD as the optimizer for local updates with a learning rate of 0.001. The local batch size is 32, and the default local epoch is 5. Each round of local training takes several gradient steps equivalent to one epoch over the private data. We limit the maximum global learning round to 200 for the federated learning methods. The model’s performance on the validation set is monitored using accuracy, and the training process is stopped early if the validation accuracy does not improve for eight consecutive epochs. We implement FHE operation using well-designed HE GPU Libraries~\cite{yang2023implementing}. 
We use the multiplicative depth of 3 for the cryptographic parameters, the scaling factor bit digit of 40, an HE packing batch size of 4096, and a security level of 128 bits as our default HE cryptographic parameters during the evaluation.

\subsection*{Image Classification Benchmark}

\subsubsection*{Dataset}

We conduct experiments on MNIST~\cite{deng2012mnist}, Fashion-MNIST (FMNIST)~\cite{Xiao2017FashionMNISTAN}, CIFAR-10~\cite{jiang2022pgada} and Street View House Numbers (SVHN)~\cite{Goodfellow2013MultidigitNR}. 
In the MNIST and FMNIST datasets, each consists of 60,000 training images with a resolution of $28 \times 28$ pixels, while the CIFAR-10 dataset features 50,000 RGB training images, each measuring $32 \times 32$ pixels. The three datasets above include a separate set of 10,000 test images, which are utilized to assess the model's performance. 
We use SVHN for a large-scale benchmark evaluation, a digit classification benchmark dataset containing 600,000 $32 \times 32$ RGB images of printed digits (from 0 to 9) cropped from pictures of house number plates. SVHN has three sets: 70,000 training, 26,000 testing sets, and an extra set with 530,000 images that are less difficult and can be used to help with the training process. 
Additionally, we adopt quantity skew~\cite{Li2021FederatedLO} over $10$, $50$, and $100$ clients to account for non-iid FL settings. Due to the page limitation, we use 10 clients for Krum, 50 clients for Multi-Krum and 100 clients for Median algorithm. Note that the combination of aggregation algorithms and the number of clients show the similar results.

\subsubsection*{Results of Image Classification Datasets}

\begin{figure}[tb]
\centering
\includegraphics[width=\linewidth]{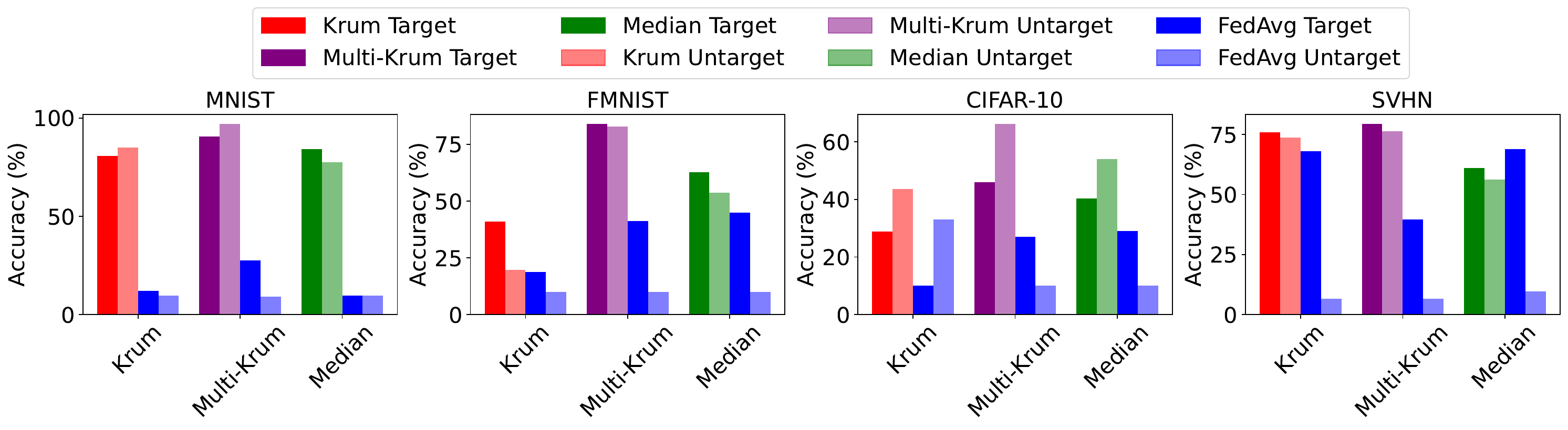 }
\vspace{-15pt}
\caption{Performance analysis of \workname~under targeted and untarget attack among Krum, Multi-Krum, Median and FedAvg across MNIST, FMNIST, CIFAR-10 and SVHN datasets. } 
\label{fig:error_analysis}
\end{figure}

We compare computation time per epoch with \workname, OpenFHE, and plaintext on MNIST, FMNIST, CIFAR-10, and SVHN datasets using LeNet-5, ResNet-18, ResNet-34, and ResNet-50 models.  Due to the page limitation and for simplicity, we only show the results of 10 clients in Krum, 50 clients in Multi-Krum, and 100 clients in Median. Note that the other results, for example 50 and 100 clients of Krum, are similar.  Firstly, we present the accuracy of Krum, Multi-Krum and Median under the targeted and untargeted attacks. As shown in Fig.~\ref{fig:error_analysis}, those three methods yield better results compared to the FedAvg. In addition, as shown in Table~\ref{tab:image_baseline}, we observe that \workname~significantly reduces the computation time while preserving information privacy. 
For example, in the Krum algorithm using the ResNet-34 model on the CIFAR-10 dataset, we  find that the time cost grows significantly, from $30.13$ seconds for plaintext up to $11,559.84$ seconds using the OpenFHE library, leading to an unacceptable time cost. However, after adopting \workname, the time reduces to $548.20$ seconds with $21\times$ speeding up from the hour level complexity to the minute level. Meanwhile, across all datasets and models, our method significantly reduces the latency compared to other works and achieves stable acceleration at constant multiplicity. We also observe the growth time is only related to the parameter size of the model and the number of clients. 

\begin{table}[tb]\small
\centering{
\begin{tabular}{ccccccc}

\toprule
\multirow{2}{*}{Dataset} & \multirow{2}{*}{Model} & \multirow{2}{*}{Model Length} & \multicolumn{3}{c}{Time (Seconds)} & \multirow{2}{*}{Speedup} \\ 
&       &     & Vanilla BRFL  & OpenFHE & \workname~(Ours)   \\
\hline
\multicolumn{7}{c}{Krum~\cite{Blanchard2017MachineLW} (10 Clients)}\\
\hline
MNIST   & LeNet-5   & 61,706      & 1.17    & 58.40    & 3.65    & $16.00\times$ \\
FMNIST	& ResNet-18 & 1,184,990  & 16.51  & 6,052.25   & 287.28  & $21.06\times$ \\
CIFAR-10	& ResNet-34 & 21,306,862 & 30.13  & 11,559.84  & 548.20  & $21.09\times$ \\
SVHN	& ResNet-50 & 23,581,695 & 33.55  & 12,746.43  & 609.20  & $20.92\times$\\
\hline
\multicolumn{7}{c}{Multi-Krum~\cite{Blanchard2017MachineLW} (50 Clients)}\\
 \hline
 MNIST   & LeNet-5   & 61,706     & 10.11	& 1,515.13	    &62.89     & $24.43\times$\\
 FMNIST	& ResNet-18 & 1,184,990  & 29.17	& 175,084.56	&6,749.32  & $25.94\times$  \\
 CIFAR-10	& ResNet-34 & 21,306,862 & 40.97 & 334,631.46	&14,185.36 & $23.59\times$  \\
 SVHN	& ResNet-50 & 11,191,262 & 44.49 & 368,748.73	&15,617.53 & $23.61\times$  \\
\hline
\multicolumn{7}{c}{Median~\cite{yin2018byzantine} (100 Clients)}\\
\hline
MNIST   & LeNet-5   & 61,706     & 19.53	    & 4905.57       &194.22    &  $25.28\times$ \\
FMNIST	& ResNet-18 & 1,184,990  & 58.17	    & 498,687.98	&23,340.56 & $21.36\times$ \\
CIFAR-10	& ResNet-34 & 21,306,862 & 80.97     & 952,216.36    &44,623.68 & $21.33\times$  \\
SVHN	& ResNet-50 & 23,581,695 & 102.49	& 1,049,547.75	&49,656.23 & $21.13\times$ \\
\bottomrule
\end{tabular}
}
\caption{Performance comparison among three methods across three datasets. Speedup refers to the acceleration ratio between \workname~and OpenFHE. Error denotes the error between \workname~and Vanilla BRFL.}
\label{tab:image_baseline}
\end{table}

\subsubsection*{Ablation Study}
We ablate \workname~to see how different factors affect its performance on the datasets above. Unless specified otherwise, we follow the same setting with the leading results in all models.  In particular, we divide our approaches into two categories, including cryptographic optimizations and hardware acceleration. Due to the hardware acceleration approach improvements, we adopt the hardware acceleration approach in algorithmic approaches, e.g.,  lazy relinearization (LazyRelin), and dynamic hoisting, for further comparison.  As shown in Table~\ref{table:ab-2}, we observe that cryptographic optimizations, including lazy relinearization and dynamic hoisting, reduce computation time among different datasets and models. In particular, we observe that in the FMNIST dataset of the Krum method, lazy relinearization provides up to $3.77\times$ speed up from $8.57 $ seconds without lazy relinearization to $2.28$ seconds with lazy relinearization. Meanwhile, dynamic hoisting provides a $40\%$ speed-up from  $0.44$ seconds to $0.32$ seconds. As for the hardware acceleration, for example, in the FMNIST dataset, Krum's hardware acceleration approach contributes to a significant speed-up since polynomial multiplications are data-independent and GPU-friendly operations, as shown in Table~\ref{table:ab-1}.  Generally, after implementing hardware acceleration, \workname~achieves an over 20-fold speed increase.
Detailed information on lazy relinearization, dynamic hoisting, and hardware acceleration can be referred to in the Method section.

\begin{table}[tb]\small
\centering{
\begin{tabular}{ccccccc}
\toprule
\multirow{2}{*}{Dataset} & \multirow{2}{*}{Model} & \multicolumn{4}{c}{Time (Seconds)} & \multirow{2}{*}{Speedup} \\
                         &                             & \multicolumn{1}{c}{W/ LazyRelin} & \multicolumn{1}{c}{W/o LazyRelin} & \multicolumn{1}{c}{W/ Dynamic Hoisting} & \multicolumn{1}{c}{W/o Dynamic Hoisting}\\
\hline
\multicolumn{7}{c}{Krum (10 Clients)} \\
\hline
MNIST	& LeNet-5	&0.07	&0.13	&0.32	&0.44 &$1.85\times$ $\backslash$ $1.37\times$  \\
FMNIST	&ResNet-18	&2.28	&8.57	&0.31	&0.42 &$3.75\times$ $\backslash$ $1.37\times$ \\
CIFAR-10	&ResNet-34	&4.19	&15.83	&0.31 &0.42 &$3.77\times$ $\backslash$ $1.37\times$\\
SVHN	&ResNet-50	&4.64	&17.50	&0.32	&0.44 &$3.77\times$ $\backslash$ $1.37\times$ \\
\hline
\multicolumn{7}{c}{Multi-Krum (50 Clients)}  \\
\hline
MNIST	& LeNet-5	&1.44	&3.15	&7.34	&10.32 &$2.18\times$ $\backslash$ $1.40\times$\\
FMNIST	&ResNet-18	&47.87	&197.62	&7.32	&10.19&$4.12\times$ $\backslash$ $1.38\times$\\
CIFAR-10	&ResNet-34	&97.13	&366.96	&7.32	&10.19&$3.77\times$ $\backslash$ $1.38\times$\\
SVHN	&ResNet-50	&107.56	&405.68	&7.36	&10.32&$3.77\times$ $\backslash$ $1.40\times$\\
\hline
\multicolumn{7}{c}{Median  (100 Clients)}   \\
\hline
MNIST	& LeNet-5	&6.33	&12.57	&29.11	&41.45 &$1.98\times$ $\backslash$ $1.42\times$\\
FMNIST	&ResNet-18	&209.32	&786.81	&29.01	&40.39 &$3.77\times$ $\backslash$ $1.39\times$\\
CIFAR-10	&ResNet-34	&384.68	&1,453.35	&29.11	&40.39 &$3.77\times$ $\backslash$ $1.39\times$\\
SVHN	&ResNet-50	&426.03	&1,606.81	&29.01	&41.49 &$3.76\times$ $\backslash$ $1.43\times$\\
\bottomrule
\end{tabular}
}
\caption{Ablation study of \workname~of cryptographic optimizations among three methods across three different datasets.}
\label{table:ab-2}
\end{table}

\begin{table}[tb]\small
\centering
\resizebox{.8\columnwidth}{!}{
\begin{tabular}{ccccc}
\toprule
\multirow{2}{*}{Dataset} & \multirow{2}{*}{Model} & \multicolumn{2}{c}{Time (Seconds)} & \multirow{2}{*}{Speedup} \\ &  &   W/ Hardware Acceleration & W/o Hardware Acceleration \\
\hline
\multicolumn{5}{c}{Krum (10 Clients)} \\ 
\hline 
MNIST   & Lenet-5   & 3.83  & 58.4 & $15.24\times$ \\ 
FMNIST  & ResNet-18 & 293.67    & 6052.25 & $20.61\times$ \\
CIFAR-10 & ResNet-34 & \multicolumn{1}{c}{559.95} & \multicolumn{1}{c}{11559.84} & $20.64\times$\\
SVHN    & ResNet-50 & \multicolumn{1}{c}{622.18} & \multicolumn{1}{c}{12746.43} & $20.48\times$\\ 
\hline
\multicolumn{5}{c}{Multi-Krum (50 Clients)}  \\ 
\hline 
MNIST   & Lenet-5   & 63.20	&1,515.13 & $23.97\times$              \\
FMNIST  & ResNet-18 & 6,901.94	&175,084.56 & $25.36\times$             \\
CIFAR-10 & ResNet-34 & \multicolumn{1}{c}{14,458.06} & \multicolumn{1}{c}{334,631.46} & $23.14\times$\\
SVHN    & ResNet-50 & \multicolumn{1}{c}{15,918.61} & \multicolumn{1}{c}{368,748.43} & $23.16\times$\\ 
\hline
\multicolumn{5}{c}{Median (100 Clients)}                            \\ 
\hline 
MNIST   & Lenet-5   & 207.10   & 4,905.32    & $23.69\times$           \\
FMNIST  & ResNet-18 & 23,929.43     & 498,687.98    & $20.84\times$           \\
CIFAR-10 & ResNet-34 & \multicolumn{1}{c}{45,703.64} & \multicolumn{1}{c}{952,216.36}& $20.83\times$ \\
SVHN    & ResNet-50 & \multicolumn{1}{c}{50,849.49} & \multicolumn{1}{c}{1,049,547.75} & $20.64\times$\\ 
\bottomrule
\end{tabular}
}
\caption{Ablation study of \workname~in hardware acceleration among three methods across three different datasets. Speedup refers to the acceleration ratio after adopting
the hardware acceleration.}
\label{table:ab-1}
\end{table}

\subsubsection*{Cryptographic Hyperparameter Analysis} 
\begin{figure}[t]
    \centering
    {\includegraphics[width=\linewidth]{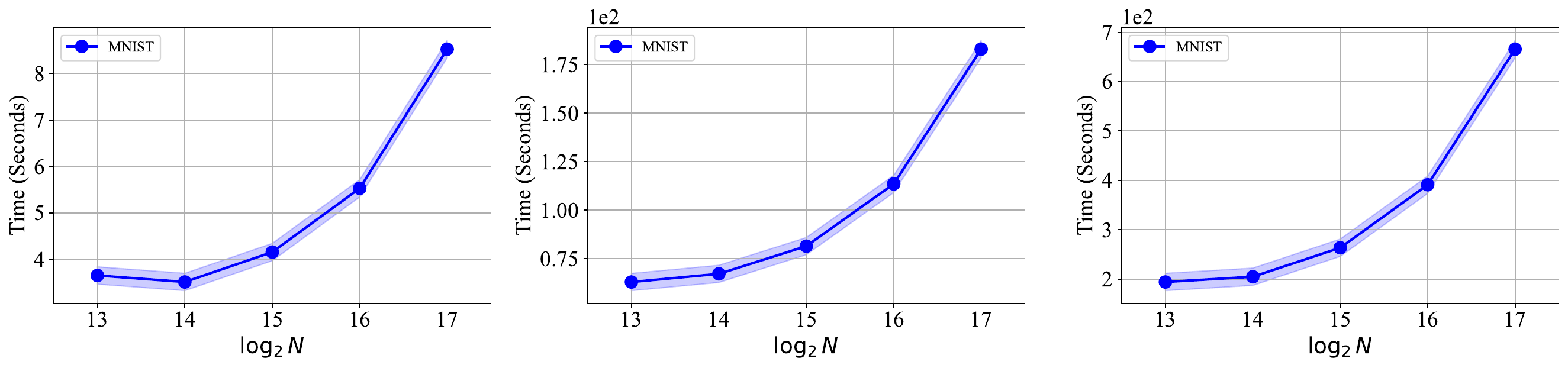}\label{fig:subfig_a}}\\
    \vspace{-20pt}
    {\includegraphics[width=\linewidth]{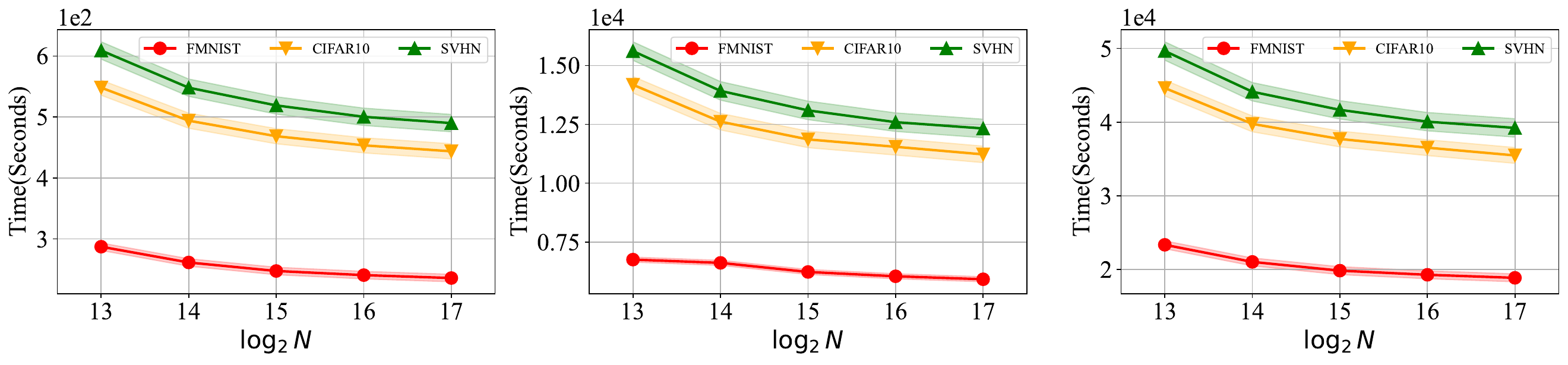}\label{fig:subfig_b}}
    \vspace{-20pt}
    \caption{FHE Packing Size $N$ Analysis. We deploy the LeNet-5,  ResNet-18, ResNet-34, and ResNet-50 models on the MNIST, FMNIST CIFAR-10, and SVHN datasets, respectively. }
    \label{fig:CHA}
\end{figure}

We provide an analysis of the impacts in FHE packing size $N$, which refers to the number of slots packed in a single ciphertext, commonly ranging from $2^{13}$ to $2^{17}$. As shown in Fig.~\ref{fig:CHA}, we first investigate the $N$ on the execution time of different models across MNIST, FMNIST, CIFAR-10, and SVHN datasets. 
A noticeable trend is the decrease in execution time for the ResNet-18, ResNet-34, and ResNet-50 models as the packing size expands, but the rate of this increase differs among models and datasets. For instance, the SVHN model on ResNet-50 sees a fairly moderate decrease in time, ranging from $609.20$ to $443.78$ seconds, as the packing size extends from $2^{13}$ to $2^{17}$. However, the LeNet-5 model on MNIST displays a converse pattern, with time initially decreasing to $3.51$ at a packing size of $2^{13}$, and then gradually decreasing to $8.53$ at $2^{17}$. 
This inconsistency may be attributed to the complexity of the models, the intrinsic characteristics of the datasets, and the efficiency of the FHE packing process.  For instance, a potential reason for why the computation time per epoch for deeper models, such as ResNet-34, decreases as the FHE packing size increases could be related to the increased capacity of a single ciphertext to hold more parameters. As the overall number of parameters remains constant and the quantity of ciphertext reduces, our hardware acceleration approach can accelerate the computation of ciphertext more efficiently, leading to a decrease in the overall computation time.

\subsection*{Medical Image Diagnosing Benchmark}
We evaluate \workname~on several biomedical tasks such as colorectal cancer, peripheral blood, thorax diseases with X-ray, etc. Note that we use the same experiment setting with the image classification benchmark and the distribution of patches in each split is described in Fig.~\ref{fig:example_dataset}.

\subsubsection*{Dataset}

We use MedMNIST~\cite{Yang2020MedMNISTCD,Yang2021MedMNISTV} to evaluate \workname. This dataset encompasses a broad range of primary data modalities including, but not limited to, X-ray, Optical Coherence Tomography (OCT), ultrasound, Computed Tomography (CT), and electron microscope images, including PathMNIST, ChestMNIST, DermaMNIST, OCTMNIST, PneumoniaMNIST, RetinaMNIST, BreastMNIST, BloodMNIST, TissueMNIST, OrganAMNIST, OrganCMNIST, OrganSMNIST datasets\footnote{Due to the page limitation, we only show the results from the aforementioned 12 datasets while \workname~can also easy adopted to other datasets in MedMNIST. Detailed information on such datasets can be referred to our Appendix.}. These datasets are used for disease diagnosis. For instance, PathMNIST can be used to assist in diagnosing colorectal cancer which is a common disease with a variable clinical course. 
The ChestMNIST, a derivative of the NIH-ChestXray dataset, comprises 112,120 frontal-view X-ray images sourced from 30,805 unique patients. Accompanying these images are 14 disease labels, text-mined to enable the classification and localization of common thorax diseases. 
To prevent data leakage, in this paper, we follow the official data split and use the same training, validation, and test set in MedMNIST~\cite{Yang2021MedMNISTV}, and for the FL setting, we adopt the same partition.



\subsubsection*{Results of Medical Diagnosing Benchmark}
Table~\ref{table:med} provides a comparison of computation time per epoch for three different methods: \workname~(Ours), OpenFHE, and vanilla BRFL. The comparison is conducted over twelve datasets in MedMNIST using various models and under different scenarios (e.g., Krum with 10 clients, Multi-Krum with 50 clients, and Median with 100 clients). Each row represents a different dataset with its particular characteristics, such as data modality, tasks (classes), the number of samples, and the implemented model. For each dataset, the computation times achieved by each of the three methods are recorded. Table~\ref{table:med} shows that \workname~significantly reduces computation time while preserving the privacy of medical data across all scenarios. For instance, in the case of Krum with 10 clients, when using the ResNet-34 model on the DermaMNIST dataset, the computation time increases significantly from 26.48 seconds in plaintext to an overwhelming 11545.84 seconds with OpenFHE. However, with \workname, the time decreases to 545.12 seconds, which is a 21-fold speed improvement, reducing the complexity from the magnitude of hours to minutes. Across all datasets and models, \workname~consistently reduces latency compared to other methods and achieves a stable acceleration at a constant multiplicity. The growth in time is only related to the parameter size of the model and the number of clients. This analysis signifies the efficiency of \workname~in reducing computation time, making it a promising method for robustly aggregating the model with privacy considerations.

\begin{table}[tb]\small
\centering
\resizebox{\columnwidth}{!}{
\begin{tabular}{cccccccc}
\toprule
\multirow{2}{*}{Dataset} & \multirow{2}{*}{Data Modality} & \multirow{2}{*}{Model} & \multirow{2}{*}{Model Length}  & \multicolumn{3}{c}{Time (Seconds)}&\multirow{2}{*}{Speedup}  \\
&                                &                                  &                                                 & Vanilla BRFL                     & OpenFHE                & \workname~(Ours)  \\ \hline
\multicolumn{8}{c}{Kurm  (10 Clients)}   \\
\hline
PathMNIST  & Colon Pathology     & LeNet-5  & 62,006   & 1.44 & 55.3 & 3.49 & $15.84\times$  \\
ChestMNIST & Chest X-ray      & ResNet-18  & 1,184,990   & 0.83 & 6040.25 & 270.43& $22.33\times$  \\
DermaMNIST & Dermatoscope      & ResNet-34  & 21,305,323  & 26.48 & 11,545.84 & 545.12& $27.82\times$ \\
OCTMNIST   & Retinal OCT         & ResNet-50  & 23,569,401  & 38.02 & 12,725.43 & 606.18 & $20.99\times$ \\
\hline
\multicolumn{8}{c}{Multi-Krum  (50 Clients)}   \\
\hline
PneumoniaMNIST & Chest X-Ray  & LeNet-5 & 61,706 & 10.96 & 1501.12 & 54.26&$22.33\times$ \\
RetinaMNIST & Fundus Camera   & ResNet-18  & 11,188,697  & 12.95  & 174,334.54 & 7,347.45&$23.72\times$ \\
BreastMNIST & Breast Ultrasound           & ResNet-34   & 21,302,758  & 16.34 & 333,731.32& 14,031.59&$23.78\times$ \\
BloodMNIST  & \makecell[c]{Blood Cell \\ Microscope}    & ResNet-50 & 23,577,597   & 38.18 & 367,623.93 & 15,448.71&$23.79\times$ \\
\hline
\multicolumn{8}{c}{Median  (100 Clients)} \\
\hline
TissueMNIST & \makecell[c]{Kidney Cortex \\ Microscope}   & LeNet-5 & 62,006  & 20.54 & 4,885.25 & 177.66&$27.49\times$\\
OrganAMNIST & Abdominal CT  & ResNet-18  & 11,191,775 & 32.35  & 498,584.98 & 23,422.38&$21.28\times$ \\
OrganCMNIST & Abdominal CT  & ResNet-34  & 21,307,375  & 51.12  & 952,066.36 & 46,720.43&$20.37\times$ \\
OrganSMNIST & Abdominal CT   & ResNet-50   & 23,583,744  & 60.53  & 1,049,337.97 & 51,982.89&$20.18\times$ \\
\bottomrule
\end{tabular}
}
\caption{Time comparison (per epoch) among three methods across 12 datasets in MedMNIST.}
\label{table:med}
\end{table}

\subsubsection*{Performance and Security Analysis}

\begin{figure}[t]
\centering
\includegraphics[width=\linewidth]{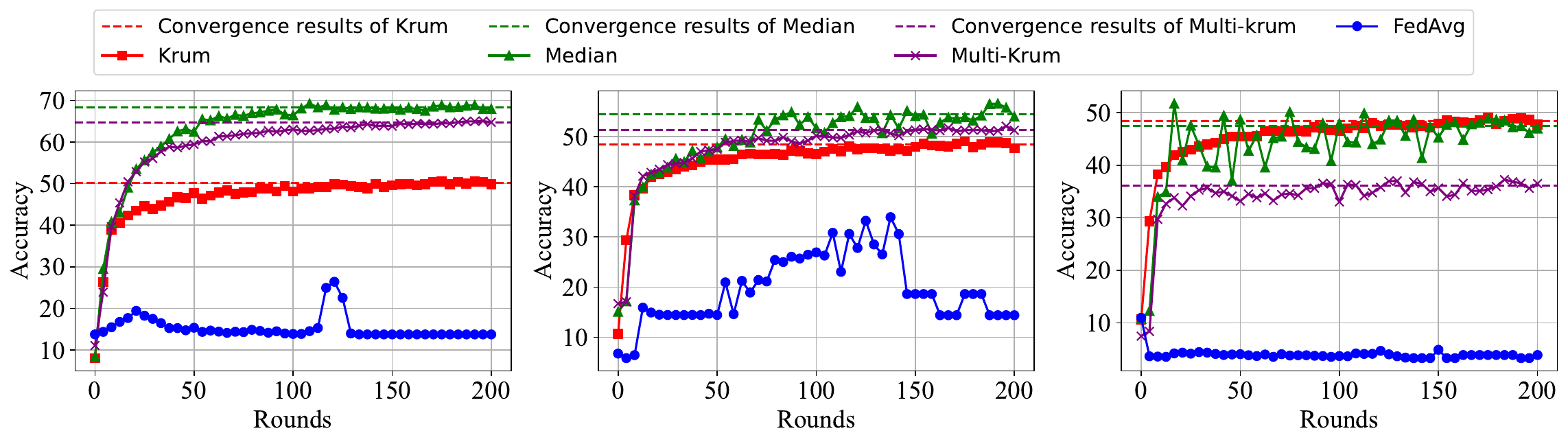}
\vspace{-15pt}
\caption{Performance analysis of \workname~under internal attack. We present the learning curves for the Krum, Median, and Multi-Krum algorithms across three datasets e.g., Bloodmnist, Pathmnist, Dermamnist in ciphertext within \workname. The horizontal lines depict the convergence outcomes of the three methods when applied to plaintext data.} 
\label{fig:miwenlubang}
\end{figure}

Here, we offer a performance analysis of \workname, considering the security levels from both internal and external attack perspectives.
Firstly, \workname~can defend against the third attack, i.e., label flipping, from malicious clients. To prove this, we conduct an empirical evaluation of several robust algorithms, employing the \workname~defense mechanism, under adversarial conditions with a Byzantine client ratio of 0.1 across three randomly selected MedMNIST datasets, e.g., Bloodmnist, Pathmnist Dermamnist. As shown in Fig.~\ref{fig:miwenlubang}, the findings reveal that our implementations of Krum, Median, and Multi-Krum algorithms, when integrated with Lancelot, yield convergence performance on par with that of non-secure, plaintext methods, even in the presence of Byzantine attacks. Notably, our approach demonstrates superior accuracy relative to the FedAvg~\cite{mcmahan2017communication} algorithm under similar Byzantine conditions, underscoring that our methodology does not compromise on robustness or inferential precision.

\begin{figure}[tb]
\centering
\includegraphics[width=\linewidth]{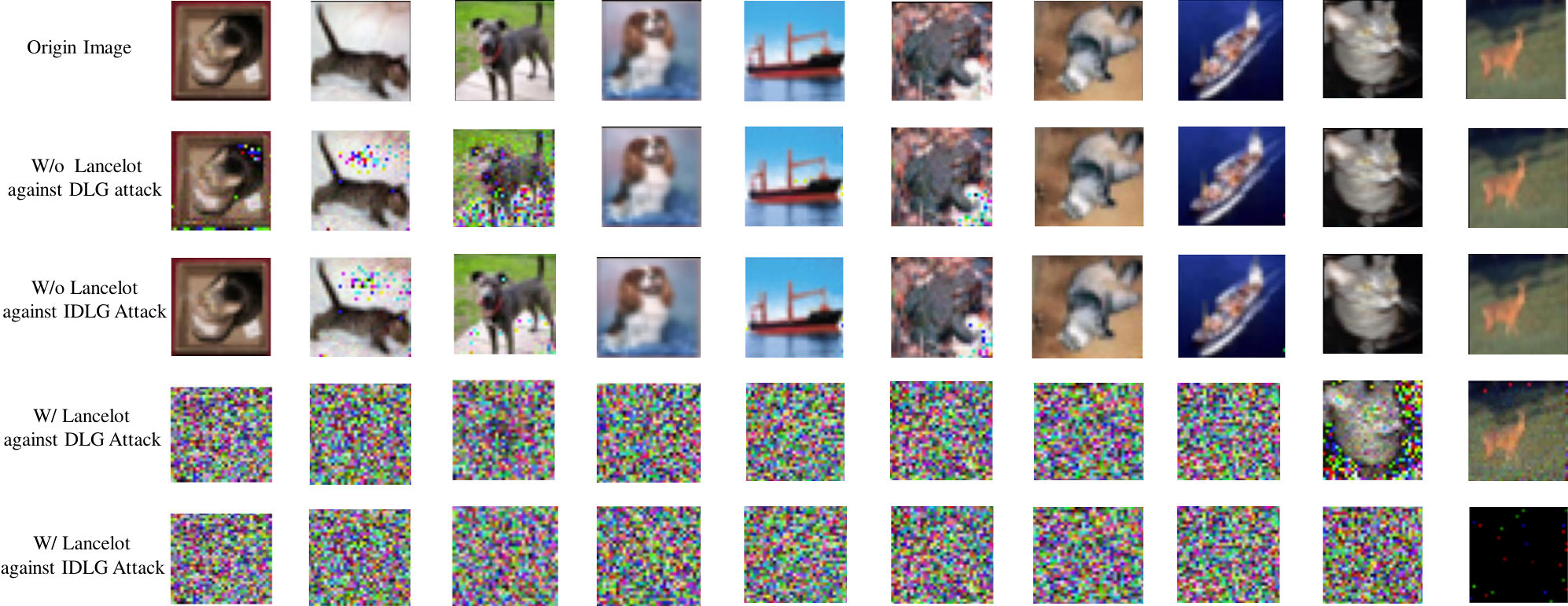}
\caption{Performance analysis of \workname under external attack. Ten pictures of the reconstruction of 32 × 32 CIFAR-10 images over the first 100 images from the validation set using the LeNet-5 as the backbone model and conservatively executing the DLG and IDLG attack for 100 iterations.}
\label{fig:dijiubu}
\end{figure}

Secondly, \workname~can also defend against the external attack. In particular, malicious clients are in possession of the public key ($\pk$) and have access solely to their own data, ensuring the safety of other clients' information. The server, equipped with the evaluation key ($\evk$), is capable of performing homomorphic computations and receives encrypted models from the clients. \workname, however, prevents the server from decrypting these models and extracting any data. Although there is communication between the server and the key generation center to implement the Byzantine-robust aggregation rules, it is important to note that all transmissions are secured with homomorphic encryption. Consequently, from Steps 1 to 8 depicted in Fig.~\ref{fig:Lacelot}, all information remains encrypted. It is only at Step 9, when the aggregated model is decrypted, that there is a potential risk of privacy breaches. However, empirical evidence demonstrates that \workname~can effectively safeguard privacy even for decrypted information. In particular, we use Deep Learning Gradient (DLG)\cite{zhu2019deep} and Iterative Deep Learning Gradient (IDLG)\cite{zhao2020idlg} to attack the FL system. As shown in Fig.~\ref{fig:dijiubu}, such attacks aim to reconstruct the images utilizing their gradient knowledge, both with and without \workname. We observe that our proposed method ensures data protection without information leakage, as evidenced by the failure to reconstruct the original image. Furthermore, the ownership of the image is hidden under the \workname~scheme. However, DLG and IDLG attacks without protection can provide recognizable images and work particularly well in a realistic setting.

To assess the performance of \workname~under DLG and IDLG attacks, we measure the Cumulative Distribution Functions (CDF) of Mean Square Error (MSE),  Structural Similarity Index Measure (SSIM) and Peak Signal-to-Noise Ratio (PSNR). Specifically, MSE quantifies pixel-wise differences, but it does not reflect the perceptual quality of an image. Conversely, SSIM and PSNR serve as metrics for assessing the decline in quality of the reconstructed image, addressing the limitations inherent in MSE, albeit not entirely aligning with human perceptual nuances. When implementing our method, we observe that most of the MSE values are concentrated on the right side of the image. This indicates that these values are generally high, suggesting that the original information has not been restored.  As an example, the details of the image (Row 4, Column 9), characterized by an MSE value of 36.66, have been reinstated. The corresponding cumulative distribution function (CDF) stands at 0.13, signifying the infrequency with which partial information restoration can occur. However, in the same image without \workname~protection, the MSE is less than $0.001$ with a CDF of $0.63$. These samples show that \workname~is effective in counteracting the inversion attack.

\begin{figure}[!ht]
  \begin{tabular}{@{}c@{}}
    \includegraphics[width=\textwidth]{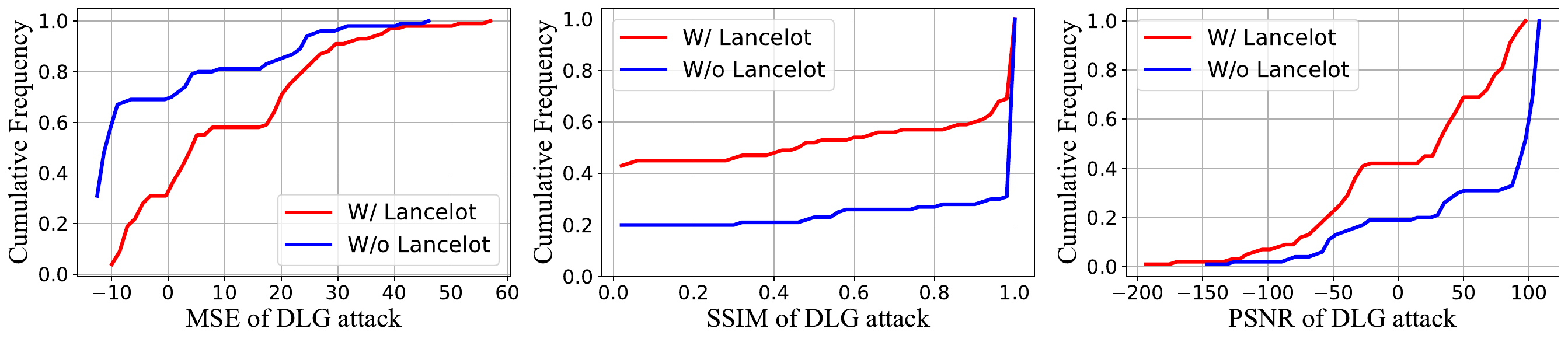} \\[\abovecaptionskip]
  \end{tabular}
  \begin{tabular}{@{}c@{}}
    \includegraphics[width=\textwidth]{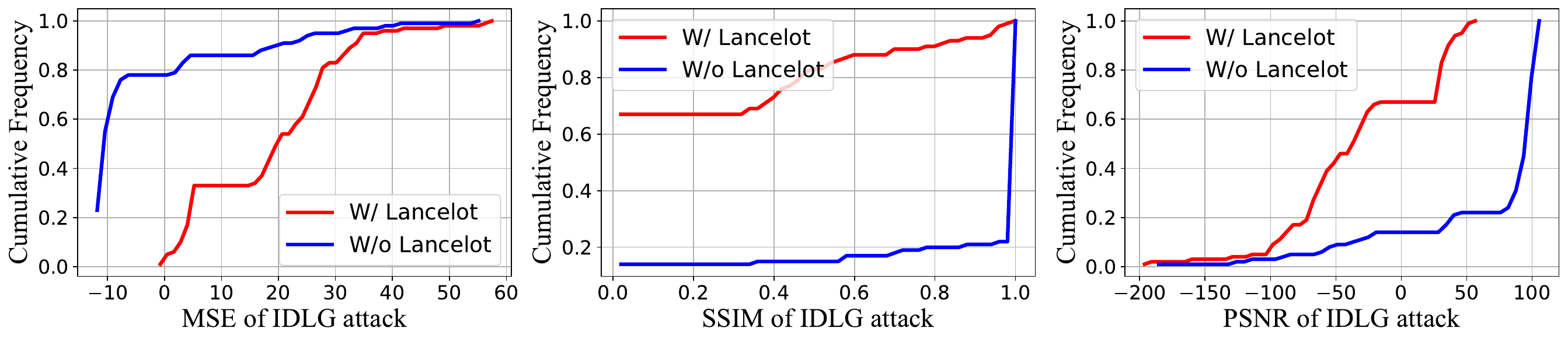} \\[\abovecaptionskip]
  \end{tabular}
    \vspace{-20pt}
    \captionof{figure}{Comparison of the Cumulative Distribution Functions (CDF) in MSE, SSIM and PSNR.}
    \label{fig:subfigures}
\end{figure}

\section*{Discussion}
FHE has recently attracted much attention in the privacy-preserving Machine-learning-as-a-service (MLaaS) application. In this paper, we demonstrate \workname, a computation-efficient privacy-preserving byzantine robust federated learning framework for several complex tasks, including image recognition and biomedical diagnosis. Our solution shows highly promising results for enhancing privacy-preserving federated learning services in a computation-effective and reliable manner, which can have abundant social and health value. Meanwhile, \workname~represents a novel approach to privacy-preserving BRFL, which combines the strengths of both conventional federated learning approaches and cryptographic frameworks for secure computation. While FHE leverages ciphertexts and increases memory usage, \workname~efficiently mitigates the computational burdens commonly associated with cryptographic solutions~\cite{froelicher2020drynx,lu2020web,kim2023secure}.  


We are confident that our proposed method has scalable potential in assisting diagnosis and various other applications, such as stroke rehabilitation and Alzheimer's disease monitoring~\cite{shuai2022balancefl,ouyang2022clusterfl}. Our study was enabled by the synergistic combination of machine-learning technologies and cryptographic development. Although significant progress in the theory and practice of FHE has been made toward improving efficiency in recent years, FHE-based approaches are believed to have a crucial bottleneck to achieving practical performance, and the cryptographic protocol is still regarded as theoretical. Recently, FHE-based machine-learning approaches\cite{kim2020semi,yang2023xnet} have demonstrated the feasibility and scalability of privacy-preserving medical data analysis. We hope our study can provide a reference for developing FHE-based secure approaches. 

Our work could be extended in several future directions to facilitate the adoption of \workname. The first direction is that large-scale deployment of \workname~would be a challenging yet important milestone for this endeavor, especially in cross-device FL scenarios.  Our work demonstrates \workname's applicability on a reliable baseline. It constitutes an essential step towards building trust in our technology and fostering its adoption, thus enabling its use to discover new scientific insights.  Next, we will extend the capabilities of \workname~by developing additional protocols for a broader range of standard analysis tools and ML algorithms in biomedical research. A key step in this direction is to make our implementation of \workname~easily configurable by practitioners for their applications. Specifically, integrating \workname~to existing user-friendly platforms such as NVFLARE~\cite{Roth2022NVIDIAFF} or FedML~\cite{he2020fedml} to make it widely available would help empower the increasing efforts to scientific discoveries. Lastly, by leveraging the FHE, the ciphertexts expand and cost more memory, but it can further optimize the communication cost from the cryptographic perspective~\cite{froelicher2020drynx,lu2020web,kim2023secure}. 
Therefore, \workname~provides an initial foundation to establish a secure and privacy-preserving ecosystem between algorithm developers and data owners.

\section*{Methods}
\workname~aims to address the two aforementioned challenges in the FHE-based BRFL system. Firstly, to mitigate the constraints imposed by the limited multiplication depth in FHE, we introduce a masked-based encrypted sorting method, which leverages interactive protocols to perform sorting operations efficiently and securely rather than relying on direct encrypted sorting.
Secondly, to reduce computational overhead, we employ a combination of two cryptographic optimizations alongside hardware acceleration, both of which synergize to expedite the entire training process. The organization of this section is shown as follows. We first outline the preliminaries of \workname. Next, we illustrate the implementation of masked-based encrypted sorting. Finally, we introduce the specific cryptographic optimizations and hardware acceleration techniques utilized in our system.


\subsection*{Problem Formulation and Threat Model}

In this work, we consider a federated learning scenario that considers a condition where $c$ out of $n$ total clients are malicious. The malicious clients may keep their toxic data but can not access the local data of other honest clients. Both honest and malicious clients can get the decrypted global model, yet local model updates uploaded by a single honest client cannot be observed. Note that 
malicious clients can launch either targeted attacks or untargeted attacks. Malicious clients can redirect the global model in the wrong direction, impairing the efforts of honest clients. 
In addition, due to the inefficient computation of HE-based  for the training phase, the goal of \workname~is to minimize the training cost, i.e., training time, and keep the same accuracy performance as the original model, which can be formulated as follows.

$$
      \min T_{total} = \sum^{E_{max}} T_{E}; \quad \quad 
      s.t. \lVert W^* - W \rVert < \delta,
$$
where $T_{total}$ is the overall training time. $W^*$ denotes the final model using the CKKS scheme for protecting. $W$ is the final aggregated model using plaintexts. $T_{E}$ is the training time for each epoch, and $E_{max}$ is the maximum epoch in training.

The threat model of this work is listed as follows.  Key Generation Center acts as a trusted third party. The server is defined as an honest but curious entity with high computation capability. This characterization implies that while the server faithfully executes established protocols, it may be interested in extracting sensitive information. Our analysis further bifurcates clients into two distinct categories. The first category comprises honest clients who endeavor to enhance the global model. They do so by uploading genuine gradients, which have been trained on their local datasets. Conversely, the second category encapsulates malicious clients. These entities deliberately upload gradients designed to degrade the accuracy of the global model. Their intent is to sabotage the effectiveness of the model, thereby posing a significant challenge in maintaining model integrity. The potential threats caused by the above entities are shown as follows. 

The following two threats are from the semi-honest server, i.e., 
\textit{Data Leakage} and  \textit{Inference Attacks.} The gradient essentially acts as a map of the client's local data. Suppose that clients directly upload these plaintext gradients; the clients inadvertently expose pathways for attackers to infer or extract the original data of honest clients, which leads to substantial data leakage, posing a significant risk to client privacy. Meanwhile, in \workname, the server and the key generation center exchange certain intermediate results to facilitate the aggregation of local updates. While essential for operation, this process could potentially be exploited to infer sensitive information from these intermediate results, posing a risk to privacy in FL.

In addition, the malicious clients also bring the threats, i.e., \textit{Poisoning Attacks.}  Malicious clients aim to manipulate the global model's performance without detection. They can execute poisoning attacks in numerous ways. Notably, through a label-flipping attack, a malicious client can alter data labels and subsequently upload gradients trained on this manipulated data. These tactics pose significant threats to the integrity and reliability of the learning process in FL.

\subsection*{Preliminaries}

\subsubsection*{Private Distance Evaluation}
In our solution, two transmissions require protection to ensure data privacy. 
The first is from the client to the server, and the second is from the server to the KGC. 
Given that the server operates under a semi-honest model and the KGC is a trusted center, we have devised distinct approaches tailored to their specific characteristics. 
This ensures both efficiency and security. The details are outlined below.

During the computation of distance, each client $\mathcal{C}_i$ sends his models $\mathbf{W}_i$ to the server, and the server computes the distance through $\mathbf{d}_{i,j} := ||\mathbf{W}_i - \mathbf{W}_j||^2$.
To protect the model information, the weight matrix should be encrypted, and we utilize a homomorphic encryption scheme to enable private distance evaluation on the server side.
In detail, we apply CKKS, a ring-variant of HE scheme that allows batching, to enable efficient processing of several computations simultaneously in a single-instruction-multiple-data manner.
The ciphertext space is defined by ring $R := \mathbb{Z}_Q[X]/(X^N+1)$, where $N$ is a power of 2 and $Q$ is the ciphertext modulus consisting of a chain of prime integers.
Let $Q := \Pi_{i=0}^L q_i$. Thus, the ciphertext is present under the residue number system (RNS) representation of $q_i$ for efficient computation.
Denote $(\pk, \sk)$ as the public and secret key pair, $\llbracket \cdot \rrbracket$ as the encryption function, and $\mathsf{Dec}_\sk$ as the decryption under $\sk$, for plaintexts $A$ and $B$, the following properties are leveraged in HE schemes:
\begin{itemize}
    \item \textbf{Homomorphic Addition / Subtraction ($\mathsf{HAdd}/\mathsf{HSub}, \oplus/\ominus$):}
        $\mathsf{Dec}_\sk(\llbracket A \rrbracket \oplus \llbracket B \rrbracket) = A+B$, 
        $\mathsf{Dec}_\sk(\llbracket A \rrbracket \ominus \llbracket B \rrbracket) = A-B$
    \item \textbf{Homomorphic Multiplication /  Square ($\mathsf{HMult}/\mathsf{HSquare}, \otimes$):}
        $\mathsf{Dec}_\sk(\llbracket A \rrbracket \otimes \llbracket B \rrbracket) = A \times B$,
        $\mathsf{Dec}_\sk(\mathsf{HSquare}(\llbracket A \rrbracket)) := A^2$
    \item \textbf{Homomorphic Rotation ($\mathsf{HRot}_k$):}
        $\mathsf{Dec}_\sk(\mathsf{HRot}_k(\llbracket A \rrbracket)) = A \gg k$
\end{itemize}
With this scheme, the new computational pattern become $\llbracket \mathbf{d}_{i,j} \rrbracket := \mathsf{HSquare}(\mathsf{HSub}(\llbracket \mathbf{W}_i \rrbracket, \llbracket \mathbf{W}_j \rrbracket))$.
The evaluation is performed on ciphertext, and only the entities with the knowledge of $\sk$ can obtain the results.

\subsubsection*{CKKS Scheme Description}

The CKKS scheme is a prominent word-wise FHE scheme that supports the approximate evaluation of real numbers at a preset precision.
This scheme allows batched computation by pre-splitting the input into different slots and is efficient on some time-consuming operations.
The security of CKKS is based on the hardness of the Ring-Learning with Errors (R-LWE) problem.
Denote the quotient ring as $R=\mathbb{Z}[x]/(x^{N}+1)$, where $N$ is a power of $2$, the R-LWE problem is defined as to distinguish the samples $(\bm{a}_i, \bm{b}_i)$ from the uniform distribution on $R_q^2$, where $\bm{b}_i = \bm{a}_i \bm{s} + \bm{e}_i$, $\bm{a}_i \stackrel{\$}{\leftarrow} R_q$, $\bm{e}_i \leftarrow \mathcal{X}$.
The security assumption of R-LWE states that there is no efficient algorithm that can distinguish the samples with non-negligible probability, forming a security assurance of the entire encryption system.

Let $\lambda$ denote the security parameter.
Let $\bm{s}$ be a random element in $R_q$ and $\mathcal{X} = \mathcal{X}(\lambda)$ be a distribution over $R_q$.
Let $Q=\prod_{i=0}^{L}q_i$ be the ciphertext modulus and $Q_\ell=\prod_i^{\ell}q_i$ be the modulus for the ciphertext at level $l$, where $q_i$ are primes.
The special modulus \cite{gentry2012fully} is denoted as $p$.
Denote $[\bm{a}]_Q$ as $a \bmod Q$.
We define the main components of the CKKS scheme as follows.

\begin{itemize}
	\item Key generation. Given the system parameter $\texttt{params}=(\lambda, N, Q)$ and distributions $\mathcal{X}_{\text{key}}$ and $\mathcal{X}_{\text{err}}$, generates the public key and secret key in the following way:
	\begin{itemize}
	    \item Secret key. Sample $\bm{s}\leftarrow \mathcal{X}_{\text{key}}$, and set the secret key $\texttt{sk} := (1, \bm{s})$. 
	    \item Public key. Sample $\bm{a}\mathop{\leftarrow}\limits^{\$}R_{Q}$ and $\bm{e}\leftarrow \mathcal{X}_{\text{err}}$. The public key is formed as $\texttt{pk} := ([-\bm{a}\cdot\bm{s}+\bm{e}]_{Q}, \bm{a})$S. 
	    \item Evaluation key. Sample $\bm{a}'\stackrel{\$}{\leftarrow} R_{p\cdot Q}$ and $\bm{e}'\leftarrow \mathcal{X}_{\text{err}}$. The evaluation key is defined as $\texttt{evk} := ([-\bm{a}'\cdot\bm{s}+\bm{e}'+p\bm{s}']_{p\cdot Q}, \bm{a}')$, where $\bm{s}'=[\bm{s}^{2}]_{Q}$. 
	    \end{itemize}
	\item Encryption. Given a public key $\texttt{pk} = (\bm{u}_{0}, \bm{u}_{1})\in R_{Q}^{2}$ and a message $m\in R$, sample $\bm{r}\leftarrow \mathcal{X}_{\text{key}}$ and $\bm{e}_{0},\bm{e}_{1}\leftarrow \mathcal{X}_{\text{err}}$. The resulting ciphertext of CKKS encryption is described as follows:
            $$\texttt{CKKS.Enc}(\texttt{pk}, m) = ([m+\bm{r}\cdot\bm{u}_{0}+\bm{e}_{0}]_{Q}, [\bm{r}\cdot\bm{u}_{1}+\bm{e}_{1}]_{Q})$$

	\item Decryption. Given a secret key $\texttt{sk} = (1, \bm{s})$ and a ciphertext $\texttt{ct} = (\bm{c}_{0}, \bm{c}_{1})\in R_{Q_\ell}^{2}$, the decryption algorithm is like the following: 
	       $$\texttt{CKKS.Dec}(\texttt{sk}, \texttt{ct}) = [\bm{c}_{0}+\bm{c}_{1}\cdot \bm{s}]_{Q_\ell}$$

	\item Evaluation. Homomorphic encryption provides a way to perform operations on ciphertext without decryption. Common homomorphic evaluations include homomorphic addition and multiplication. For the three schemes, we give their detailed evaluation procedures below.
	    \begin{itemize}
	    \item CKKS Addition. Given two CKKS ciphertexts $\texttt{ct}$ and $\texttt{ct}'$ in $R_{Q_\ell}^{2}$, their sum is defined  as the follows:
            $$\texttt{CKKS.Add}(\texttt{ct}, \texttt{ct}') = [\texttt{ct} + \texttt{ct}']_{Q_\ell}$$
	    
	    \item CKKS Multiplication. Given two CKKS ciphertexts $\texttt{ct} = (\bm{c}_{0}, \bm{c}_{1})$, $\texttt{ct}' = (\bm{c}_{0}', \bm{c}_{1}')$ in $R_{Q_\ell}^{2}$, the product of which yields a triple defined as the follows:
            $$\texttt{CKKS.Mult}(\texttt{ct}, \texttt{ct}') = (\tilde{\bm{c}}_{0}, \tilde{\bm{c}}_{1}, \tilde{\bm{c}}_{2})=[(\bm{c}_{0}\cdot\bm{c}_{0}', \bm{c}_{0}\cdot\bm{c}_{1}' + \bm{c}_{1}\cdot\bm{c}_{0}', \bm{c}_{1}\cdot\bm{c}_{1}')]_{Q_\ell}$$
	    
	    \end{itemize}
\end{itemize}

\subsubsection*{Batch Encoding and Decoding}
The plaintext space in the CKKS scheme is set to be the quotient ring $R=\mathbb{Z}[x]/(x^{N}+1)$. To enable approximate operations on floating point values, a scalar $\Delta\geq 1$ is introduced, and the encoding procedure  $\texttt{CKKS.Encode} $ is to transform vectors of complex numbers into plaintext polynomials for encryption. Given a complex primitive $2N$-th root of unity $\xi =e^{\pi i/N}$, the canonical embedding of $p(x)\in\mathbb{R}[x]/(x^{N}+1)$ into $\mathbb{C}^{N}$ is defined as 
$$
\sigma:~p(x)\longmapsto(p(\xi^{j}))_{j\in\mathbb{Z}_{2N}^{*}}. 
$$
Note that $\sigma$ is injective by Fundamental Theorem of Algebra and its image is the subring $\mathbb{H}=\{(z_j)_{j\in\mathbb{Z}_{2N}^{*}}:z_{2N-j}=\bar{z_j},\forall j\in\mathbb{Z}_{2N}^{*}\}\subseteq\mathbb{C}^{N}$. Let $T$ be a subgroup of the multiplicative group $\mathbb{Z}_{2N}^{*}$ with order $N/2$. Then the subring $\mathbb{H}$ can be identified with $\mathbb{C}^{N/2}$ via the natural projection $\pi :~(z_j)_{j\in\mathbb{Z}_{2N}^{*}}\mapsto (z_j)_{j\in T}$. To clarify, the whole transformation can be illustrated as
$$
\mathbb{R}[x]/(x^{N}+1)\mathop{\longrightarrow}\limits^{\sigma}\mathbb{H}\mathop{\longrightarrow}\limits^{\pi}\mathbb{C}^{N/2},
$$
which nearly defines the decoding procedure except for $\Delta^{-1}$:
$$\texttt{CKKS.Decode}(m\in R;~\Delta):~R\longrightarrow\mathbb{C}^{N/2},~m\longmapsto z=\pi\circ\sigma(\Delta^{-1}\cdot m)$$
The encoding is roughly the inverse of the decoding procedure with a round-off operation to make the output,t integral:
$$\texttt{CKKS.Encode}(z\in \mathbb{C}^{N/2};~\Delta):~\mathbb{C}^{N/2}\longrightarrow R,~z\longmapsto m=\lfloor\Delta\cdot\sigma^{-1}(\pi^{-1}(z))\rceil_{R}$$
Note that $\sigma:~\mathbb{R}[x]/(x^{N}+1)\rightarrow\mathbb{H}$ and $\pi:~\mathbb{H}\rightarrow\mathbb{C}^{N/2}$ are isometric ring isomorphisms, so (1) $\sigma^{-1}$ and $\pi^{-1}$ are well-defined; (2) both directions of transformations are size-preserved so that the rounding and scalar can help with precision control. 

\subsection*{Masked-based Encrypted Sorting}

In order to execute the byzantine-robust aggregation rules, the server needs to sort the clients according to the corresponding distance. However, given the constraints imposed by the multiplication depth, relying on direct sorting is impractical, especially in large FL systems with lots of clients. This is primarily because sorting algorithms operating in the ciphertext space generally depend on pairwise ciphertext comparisons, which consume multiple levels of multiplication depth. 
To address this, we propose masked-based encrypted sorting, and its core concept is to utilize an interactive protocol between the server and the key generation center that securely conveys the index information, ensuring both efficiency and privacy. 

In particular, the server sends the distance list to the key generation center. However, since the server is not allowed to know the specific index of selected clients, the key generation center needs to encrypt the index information and send it to the server, where the server can execute the aggregation rules. 
Let the encrypted distance matrix be $L$; each element $l \in L$ is encrypted distance data. The key generation center obtains the encrypted distance list, then it decrypts the encrypted distance list and obtains the sorting permutation $\pi$ of $n$ elements, which is defined as
$$
    \pi: \{1,\cdots, n\} \rightarrow \{1,\cdots, n\}.
$$

The sorting permutation $\pi$ can be represented in two-line form by
$$
\begin{pmatrix}
 1, & 2, & \cdots & n \\
 \pi(1), & \pi(2), & \cdots & \pi(n)
\end{pmatrix}.
$$

Therefore, the $n \times n$ permutation matrix $P_{\pi} = p_{ij}$ obtained by permuting the columns of the identity matrix $I_n$,  that is, for each $i$, $p_{ij}$ = 1 if $j = \pi(i)$ and $p_{ij} = 0$ otherwise. Since the entries in row $i$ are all $0$ except that a $1$ appears in column $\pi(i)$, we have

$$
P_{\pi} = 
\begin{bmatrix}
\mathbf{e}_{\pi(1)}, &
\mathbf{e}_{\pi(2)}, &
\cdots &
\mathbf{e}_{\pi(n)}
\end{bmatrix}^T,
$$
where $\mathbf{e}_j$ is a standard basis vector, denotes a row vector of length $n$ with $1$ in the $j^{th}$ position and $0$ in every other position.
Note that the key generation center knows the aggregation rules of the whole FL system. Therefore, the $P_{\pi}$ only uses the $\mathbf{e}_j$ for the selected position while the other position is $\mathbf{o}$, where $\mathbf{o}$ is a zero vector. For example, if there are 5 clients in the whole FL system, and we only use the top 3 clients, e.g., clients 1, 3, 5 for aggregation in one round, $P_{\pi} = \begin{bmatrix} \mathbf{e}_{\pi(1)},\mathbf{e}_{\pi(3)}, \mathbf{e}_{\pi(5)}, \mathbf{o}, \mathbf{o}   \end{bmatrix}$.

To protect the index information, it is also necessary to encode the permutation matrix $P_{\pi}$ and then send it back to the server. Let $\Phi$ be the encoding mapping that encodes the sorting index from the plaintext into an encrypted mask $\llbracket M\rrbracket$, which can be formulated as follows,
$$
    \llbracket M \rrbracket =  \Phi(P_{\pi}) = \{ \mathsf{Enc_{pk}}(\mathbf{e}_{j})| j \in n \}.
$$

When the server receives the encrypted mask $\llbracket M \rrbracket $, the server uses the multiplication results of the encrypted weights $\llbracket W \rrbracket$ and $\llbracket M \rrbracket $ to obtain the aggregated model based on the aggregation rules. The reason is that after adopting the mask, the $\llbracket W \rrbracket$ only transmits the information without any information leakage.


\subsection*{Lazy Relinearization}

\begin{figure}[tb]\small
  \centering{
  \includegraphics[width=.6\linewidth]{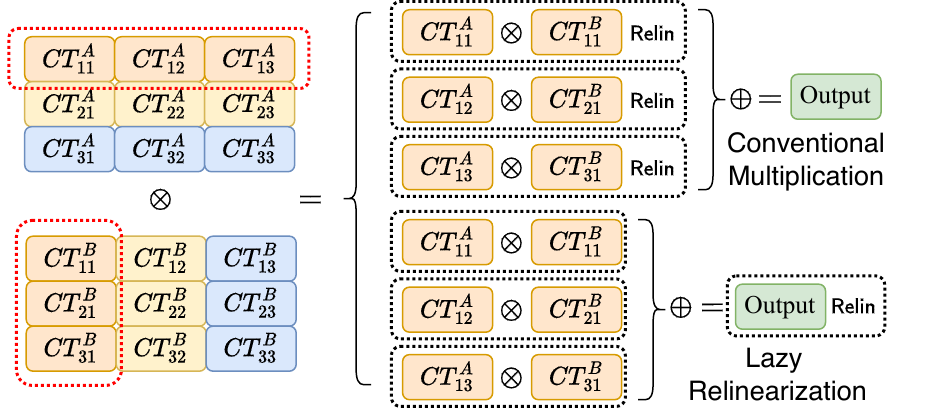}
  \caption{Illustration of Lazy Relinearization in \workname. We have two matrices, i.e., $CT^A, CT^B$, for multiplications in the ciphertext. Lazy Relinearization can reduce two relinearizations to reduce the computation time. }
  \label{fig:lazy_relin}
  }
\end{figure}

The computation of homomorphic matrix multiplication evaluation traverses the elements in the matrix recursively to perform element-by-element homomorphic multiplication and sums the relevant results.
Here, in $i$-th iteration, we take two CKKS ciphertexts $\ct^{(i)}_A := (\boldsymbol{c}^{(i)}_{A,0}, \boldsymbol{c}^{(i)}_{A,1}), \ct^{(i)}_B := (\boldsymbol{c}^{(i)}_{B, 0}, \boldsymbol{c}^{(i)}_{B, 1})$, where each consists a pair of ring elements, perform homomorphic multiplication and obtain a ring element triple $\ct^{(i)*} := (\boldsymbol{c}^{(i)*}_0,\boldsymbol{c}^{(i)*}_1, \boldsymbol{c}^{(i)*}_2)$, then return to the ring element pair $\ct^{(i)} := (\boldsymbol{c}^{(i)}_0, \boldsymbol{c}^{(i)}_1)$ through relinearization, and finally add all $\ct^{(i)}$ up.
The relinearization bottlenecks this computation, while the homomorphic addition is much cheaper as it only performs coefficient-wise addition.
Due to this consideration, we devise a lazy relinearization technique.
Denote $m$ as the number of iterations, different from the original method that computes $\sum_{i=0}^{m-1}\ct^{(i)}$, we keep the multiplication result in ternary form $\ct^{(i)*}$, and performs the sum of all ternaries directly, i.e., $\sum_{i=0}^{m-1}\ct^{(i)*}$.  
Thus, we only need to linearize the summation result, thus reducing the number of relinearization operations to one.
This technique increases the number of additions, which is far less than the overhead of relinearization, resulting in a performance boost. For example, as illustrated in Fig.~\ref{fig:lazy_relin}, we initially have three ciphertext multiplication and addition operations. In conventional multiplication, these three ciphertexts undergo multiplication and relinearization three times, followed by an add operation. However, with lazy relinearization, the relinearization step is deferred until after the addition operation. Compared to conventional multiplication, lazy relinearization eliminates two relinearization operations, resulting in significant computation time savings.

\subsection*{Dynamic Hoisting}

During the computation of the square of the matrix $\mathbf{W}$, we take a row and a column of $\mathbf{W}$ and multiply them element-wisely, and sum up all the multiplication results to obtain one element of the result vector.
Denote the dimension of $\mathbf{W}$ is $n\times n$, where $n$ is a power of 2. We need to perform homomorphic rotations to obtain the addition result.
In detail, to compute $\llbracket \mathbf{y} \rrbracket := \llbracket \mathbf{W}_t \rrbracket \otimes \llbracket \mathbf{W}^T_m \rrbracket$, where $t,m \in [0,n)$, we perform $\llbracket \mathbf{y}_{i}^{(0)} \rrbracket := \mathsf{HMult}(\llbracket \mathbf{W}_t \rrbracket, \llbracket \mathbf{W}_m \rrbracket)$, and then iterative compute $\llbracket \mathbf{y}_i^{(j+1)} \rrbracket := \mathsf{HAdd}(\llbracket \mathbf{y}_i^{(j)} \rrbracket, \mathsf{HRot}_{j}(\llbracket \mathbf{y}_i^{(j)} \rrbracket))$, thus finally obtain $\llbracket \mathbf{y} \rrbracket := \llbracket \mathbf{y}_i^{(\log n)} \rrbracket$.
Here, the sequence of successive homomorphic rotations and addition processes involves a series of automorphism and key-switching operations.
For each $\mathbf{y}_i^{(j)} = (\boldsymbol{c}_{i,0}^{(j)}, \boldsymbol{c}_{i,1}^{(j)})$, the original processing is as follows:
\begin{itemize}
    \item Automorphism: Denote the current automorphism as $\phi_k$, we first apply the automorphism to each part of $\mathbf{y}_i^{(j)}$ and computing $\mathbf{y}_i^{(j)'} = (\boldsymbol{c}_{i,0}^{(j)'}, \boldsymbol{c}_{i,1}^{(j)'}) := (\phi_k(\boldsymbol{c}_{i,0}^{(j)}), \phi_k(\boldsymbol{c}_{i,1}^{(j)}))$.
    
    \item Key-Switching: Because this process implicitly changes the key, we need to perform a key-switching operation. This contains the following process:
        \begin{itemize}
            \item RNS decompose: denote $\texttt{dnum}$ as the RNS decompose number, and $\texttt{dnum} = \lceil (L+1)/\alpha \rceil$, padding zero and then spliting $\boldsymbol{c}_{i,1}^{(j)'}$ into $\beta = \lceil (\ell+1)/\alpha \rceil$ parts, and multiplying with $[Q' := \Pi_{i=\ell +1}^{\alpha\beta -1}]_{q_{y\alpha + x}}$, where $0\leq x < \alpha$, $0\leq y < \beta$.
            \item ModUp: Raise the modulus of each part to the modulus base of the key-switching operation.
            \item Switch Key: Perform inner-product with the key-switching keys.
            \item ModDown: Converse the modulus of the results to the original $Q_\ell$.
        \end{itemize}
\end{itemize}
One shortcoming of this computation is that the homomorphic addition and rotation are performed in an iterative pattern, which makes the entire parallelism equal to the polynomial dimension $N$.
Additionally, during each interaction, we need to perform the fixed computational flow, include automorphism, raise the modulus, switch the underline key, and then bring the modulus back, which has high computation consumption.

To solve this issue, we unfold this iterative computation and get $\llbracket \mathbf{y}_i \rrbracket := \llbracket \mathbf{y}_i^{(0)} \rrbracket \oplus \sum_{k=0}^{n -1} \mathsf{HRot}_{k}(\llbracket \mathbf{y}_i^{(0)} \rrbracket)$, thus the $\log n$ times homomorphic rotation can be performed simultaneously, bringing a $\log n$-fold increase in parallelism.
Meanwhile, since this unfold approach introduces more computation, we utilize the hoisting technique \cite{DBLP:conf/crypto/HaleviS18} and reverse the order of automorphism and modulus raising to reduce computation.
The key observation of this technique is that we can reverse the order of the first three steps above without affecting the correctness of the procedure.   
In particular, this key changed computational flow is as follows:
\begin{itemize}
    \item RNS decompose and ModUp: Padding zero and splitting $\boldsymbol{c}_{i,1}^{(j)}$ into $\beta$ parts and multiplying with $[Q']_{q_{y\alpha + x}}$, raising the modulus of each part to the modulus base of key-switching operation.

    \item  Automorphism: Denote the current automorphism as $\phi_k$, applying the automorphism to each part of the above step.
\end{itemize}
Afterward, perform the original inner product with the key-switching keys corresponding to each $\phi_k$ and finally converse the ciphertext modulus to $Q_\ell$.
After all computations of rotations are computed, which can be performed in parallel, we sum up all results to obtain $\llbracket \mathbf{y}_i \rrbracket$.

\begin{figure}[t]
\centering
\includegraphics[width=.85\linewidth]{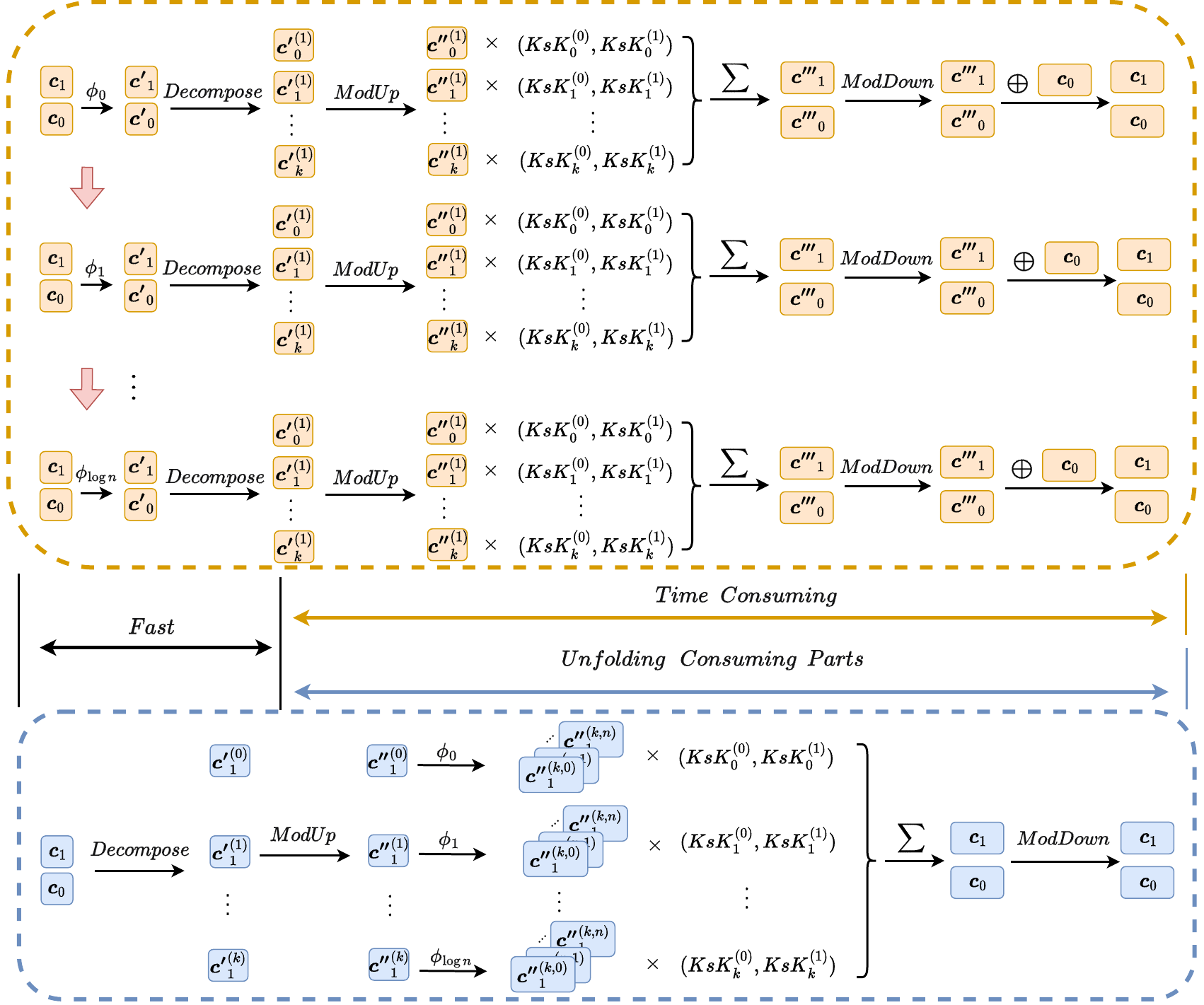}
\caption{Illustration of Dynamic Hoisting in \workname.}
\label{fig:hoisiting}
\end{figure}

\begin{figure}[!ht]
\centering
\includegraphics[width=\linewidth]{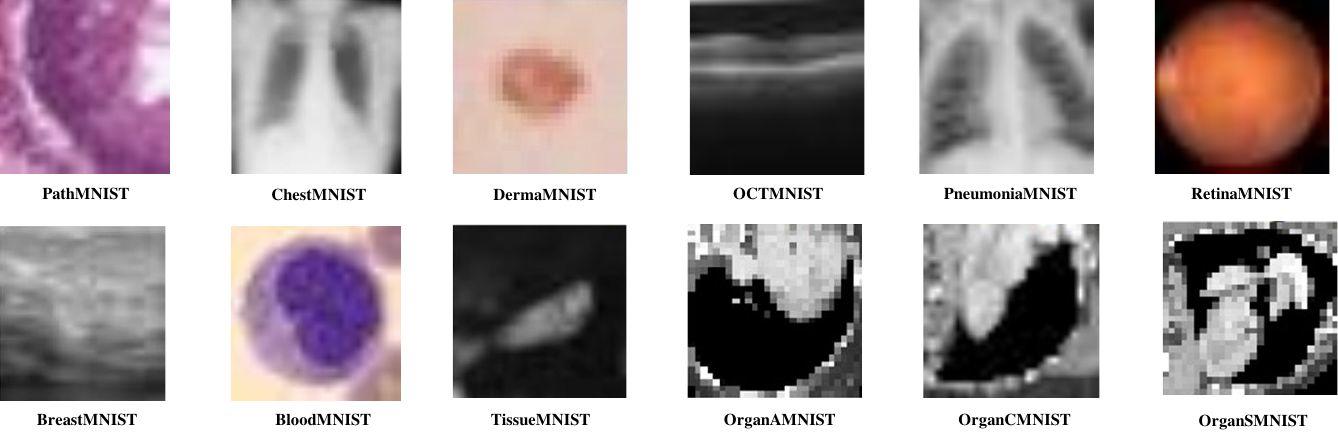}
\caption{Example of MedMNIST Dataset. We use 12 datasets, including PathMNIST, ChestMNIST, DermaMNIST, OCTMNIST, PneumoniaMNIST, RetinaMNIST, BreastMNIST, BloodMNIST, TissueMNIST, OrganAMNIST, OrganCMNIST, OrganSMNIST.}
\label{fig:example_dataset}
\end{figure}

As the automorphism distributes over addition and multiplication and will not significantly change the norm of an element, the correctness is ensured.
Through this, the time-consuming operations of modulus raising up and down are required only once. For instance, we present an illustrative example in Fig.~\ref{fig:hoisiting}.  The computation of the time-consuming components is parallelized (in the light-blue box), which significantly reduces computational overhead. In contrast, traditional methods (in the orange box) necessitate sequentially repeating the same computational procedures multiple times.

However,  when $n$ is large, this unfolding approach in the hoisting technique exponentially increases the number of rotations, significantly expanding the cipher space due to the generating of several ciphertexts simultaneously. It is not practical in real applications due to the  memory limitation in heterogeneous devices. Therefore, We devise a dynamic adjusting approach to the unfoldings to balance computation and parallelism better. In this case, instead of unfolding all the iterations, we only grow some of them, which optimizes the overall performance. We formulate this problem as follows.

$$\min (\log n - k + 1) T_{H} + (k-1) T_{D} ; \quad \quad s.t. k M_{c} \leq M_{B},
$$
where $k$ represents the unfolding parameters, while $T_{H}$ and $T_{D}$ denote the hoisting and decomposition times, respectively, which are dictated by hardware characteristics. Additionally, $M_c$ refers to the memory required by a ciphertext, and $M_B$ signifies the memory budget limit. Due to the problem being a linear convex optimization problem, we use the simplex method~\cite{nelder1965simplex} to solve it. Note that $k$ can be determined in advance before the FL training, based on the memory of the device, and $k$ is dynamically adjusted if other processes consume memory via recalculating the aforementioned optimizing problem.

\subsection*{Hardware Acceleration}
In \workname, the primary area of concern has been the server-side homomorphic evaluations of distance matrices, a computationally intense task.
This consumption mainly results from the CKKS homomorphic operations, which inherently possess a robust parallelism.
The computation of matrix elements is independent, allowing for concurrent execution of many operations without synchronization.
Thus, since servers are typically equipped with superior computational capabilities, we utilize GPU acceleration for the server-side homomorphic evaluations.
Each thread is assigned one coefficient of the ring elements for every HE operation executed and then carries out the corresponding computation, ensuring that multiple coefficients are processed simultaneously.
Furthermore, we develop a comprehensive GPU-based implementation to host the entire distance computation on GPU. 
This approach necessitates just one data interaction between the CPU and GPU, with all subsequent operations retaining their data in the GPU's global memory, significantly curtailing IO latency and boosting overall performance. The details of hardware acceleration are presented in Appendix.

\bibliography{sample}



\section*{Author contributions statement}
S. Jiang, H. Yang, Q. Xie, contributes to motivation and framework design. 
S. Jiang contributes to algorithm  masked-based encrypted sorting. H. Yang contributes to robust aggregation realization based on CKKS and cryptography knowledge. S. Jiang and Q. Xie conducted the experiment(s) and analyzed the results. C. Ma, S. Wang, G. Xing  contributes the partial drafts of this manuscript. All authors review and revise the manuscript.

\clearpage
\section*{Appendix}

\subsection*{MedMNIST Dataset Description}

MedMNIST encompasses a broad range of primary data modalities including, but not limited to, X-ray, Optical Coherence Tomography (OCT), ultrasound, Computed Tomography (CT), and electron microscope images, including PathMNIST, ChestMNIST, DermaMNIST, OCTMNIST, PneumoniaMNIST, RetinaMNIST, BreastMNIST, BloodMNIST, TissueMNIST, OrganAMNIST, OrganCMNIST, OrganSMNIST datasets In particular, PathMNIST can be used to assist in diagnosing colorectal cancer which is a common disease with a variable clinical course. PathMNIST is based on a prior study for predicting survival from colorectal cancer histology slides, providing a dataset of 100,000 non-overlapping image patches from hematoxylin and eosin stained histological images and a test dataset of 7,180 image patches from different clinical centers. The ChestMNIST, a derivative of the NIH-ChestXray dataset, comprises 112,120 frontal-view X-ray images sourced from 30,805 unique patients. Accompanying these images are 14 disease labels, text-mined to enable the classification and localization of common thorax diseases. Based on the HAM10000 dataset~\cite{tschandl2018ham10000}, the DermaMNIST is a voluminous collection of dermatoscopic images from multiple origins, primarily showcasing common pigmented skin lesions. The OCTMNIST dataset, derived from a pre-existing dataset, encapsulates 109,309 valid optical coherence tomography (OCT) images specifically curated for the study of retinal diseases. PneumoniaMNIST includes 5,856 pediatric chest X-Ray images specifically intended to classify pneumonia against normal instances. The RetinaMNIST dataset, based on the DeepDRiD challenge, furnishes a dataset of 1,600 retina fundus images. The primary task associated with this dataset is the ordinal regression for 5-level grading of diabetic retinopathy severity. The BreastMNIST dataset, derived from a collection of 780 breast ultrasound images, is designed to classify breast cancer. BloodMNIST is based on a dataset of individual normal cells obtained from individuals without any infection, hematologic, or oncologic disease and who were free of any pharmacologic treatment at the time of blood collection. The primary objective of this dataset is to recognize different types of normal peripheral blood cells. TissueMNIST leverages the BBBC05134 dataset~\cite{Woloshuk2020.06.24.167726} to classify cell types in human kidney tissue. The dataset was segmented from 3 reference tissue specimens and organized into eight categories.  To prevent data leakage, in this paper, we  follow the official data split and use the same training, validation, and test set in MedMNIST~\cite{Yang2021MedMNISTV}, and for the FL setting, we adopt the same partition.

\subsection*{Designing Objective}
We aim to design a privacy-preserving FL scheme that can resist poisoning attacks, reduce computational
overheads, and provide privacy guarantees. At the same time, \workname~should achieve the same or almost the same accuracy as the model obtained using an unencrypted training scheme. Specifically, we are committed to achieving the following goals:

\begin{itemize}
    \item  \textbf{Efficiency.} \workname~should reduce the computational overheads caused by encryption in traditional privacy-preserving FL schemes.
    \item \textbf{Scalability.} The privately robust aggregation protocol should be implemented efficiently and hence applied to large-scale FL systems that involve hundreds of parties and advanced neural network architectures. 
    \item  \textbf{Robustness.} Our scheme should be robust against malicious attacks, which means that the accuracy of the global model should not be affected by malicious clients. \item  \textbf{Accuracy.} Our scheme should \textbf{not} sacrifice accuracy while preserving privacy and resisting malicious attacks. The accuracy of the global model should be as close as possible to that trained by an unencrypted training scheme.
    \item  \textbf{Privacy.} We aim to protect the clients’ data from being compromised. Neither a third party nor the malicious parties can get or infer the original information of clients.
\end{itemize}


\subsection*{Descriptions of Hardware Acceleration}

We show the structure of the implemented framework in Figure 2 to give a concise overview. In functionality, it consists of two main parts, one serving for pre-computation and the other containing the optimized implementation of the three HE schemes that can be divided into three layers: a math/polynomial layer, an RNS arithmetic layer, and a scheme layer. The basic layer contains the low-level modular operations for both 64-bit and 128-bit
integers, a pseudo-random generator, and polynomial arithmetics. For the integer operations,
we provide well-optimized implementation written in CUDA PTX assembly to minimize
the number of machine instructions and register usage after compiling. Based on this, we implement polynomial arithmetic, such as NWT and FFT, for fast and low-complexity computation. At the middle layer, we implement the sampling and RNS arithmetic modules, of which the operands are polynomials under RNS or double-CRT representation. The sampling module consists of three approaches for sampling polynomial coefficients from ternary, uniform, and centered binomial distributions. The RNS module offers efficient polynomial arithmetics, and we implement common algorithms of both BEHZ- and HPS-type
base conversion.  The top layer offers high-level unified implementation of the BGV, BFV and CKKS schemes. For all schemes, our framework supports the following features: (1) both symmetric and asymmetric encryption; (2) homomorphic addition, subtraction and multiplication of two or multiple plaintexts and ciphertexts; (3) in-place holomorphic exponentiation, negation, and rotation of a single ciphertext.

\subsubsection*{CPU-GPU Computation Model and the architecture of GPU}

\begin{figure}[t]
\centering
\includegraphics[width=.75\linewidth]{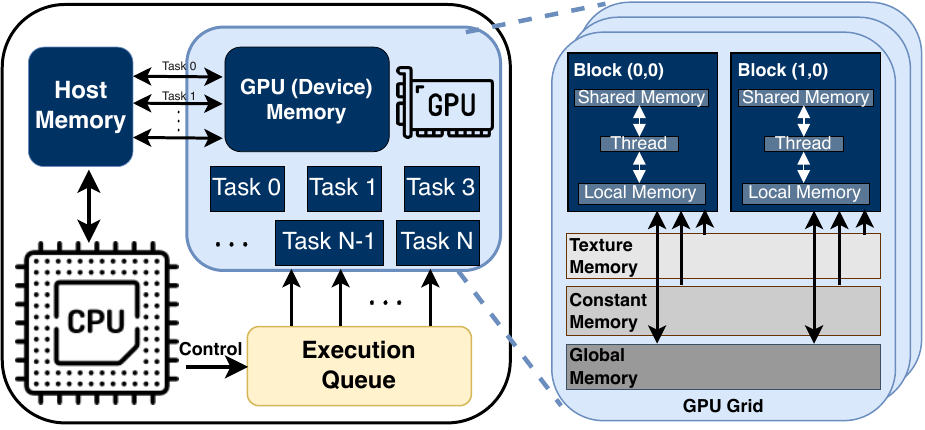}
\caption{The CPU-GPU computational model and the architecture of GPU.}
\label{fig:gpu}
\end{figure}

We present a summary of the computational model and GPU architecture in Fig.~\ref{fig:gpu}. GPU memory is categorized into two types: read-write and read-only. Read-write memory includes Global Memory (GMEM), Shared Memory (SMEM), and the Register File (RF), with access speeds ranging from slow to fast, respectively. Read-only memory encompasses Constant Memory and Texture Memory, both of which can be cached. A CUDA kernel operates on the GPU with many threads running concurrently; these threads are the smallest execution units and are organized into blocks. Each thread has its private RF, while SMEM is shared among all threads within the same block. Both GMEM and read-only memory are accessible to all threads and have the longest lifespan, covering the duration of the entire computational task. During execution, threads are grouped in warps of 32. A Streaming Multiprocessor (SM) accommodates multiple blocks, and within the SM, each Warp Scheduler (WS) manages the warps, executing them sequentially.

In a heterogeneous platform equipped with both CPU and GPU, a typical collaborative computing approach involves the CPU managing the execution queue, dispatching kernels to the GPU, and handling data transfers via PCIe before awaiting the return of results from the GPU. An alternative approach utilizes unified memory, merging CPU and GPU memory spaces to obviate the need for data transfers. Despite the ability to prefetch data, this method often falls short in performance. The former approach may lead to redundant data transfers if tasks are interdependent, while the latter suffers from subpar performance despite its prefetching capabilities. To mitigate these issues, fusing data-dependent kernels can somewhat diminish the I/O latency resulting from data transfers and memory access. However, for optimal performance, it's crucial to avoid excessive fusion that can lead to reduced SM occupancy due to high resource consumption by a block.

\subsubsection*{Details of Hardware Framework Structure}

\begin{figure}[!ht]
\centering
\includegraphics[width=.85\linewidth]{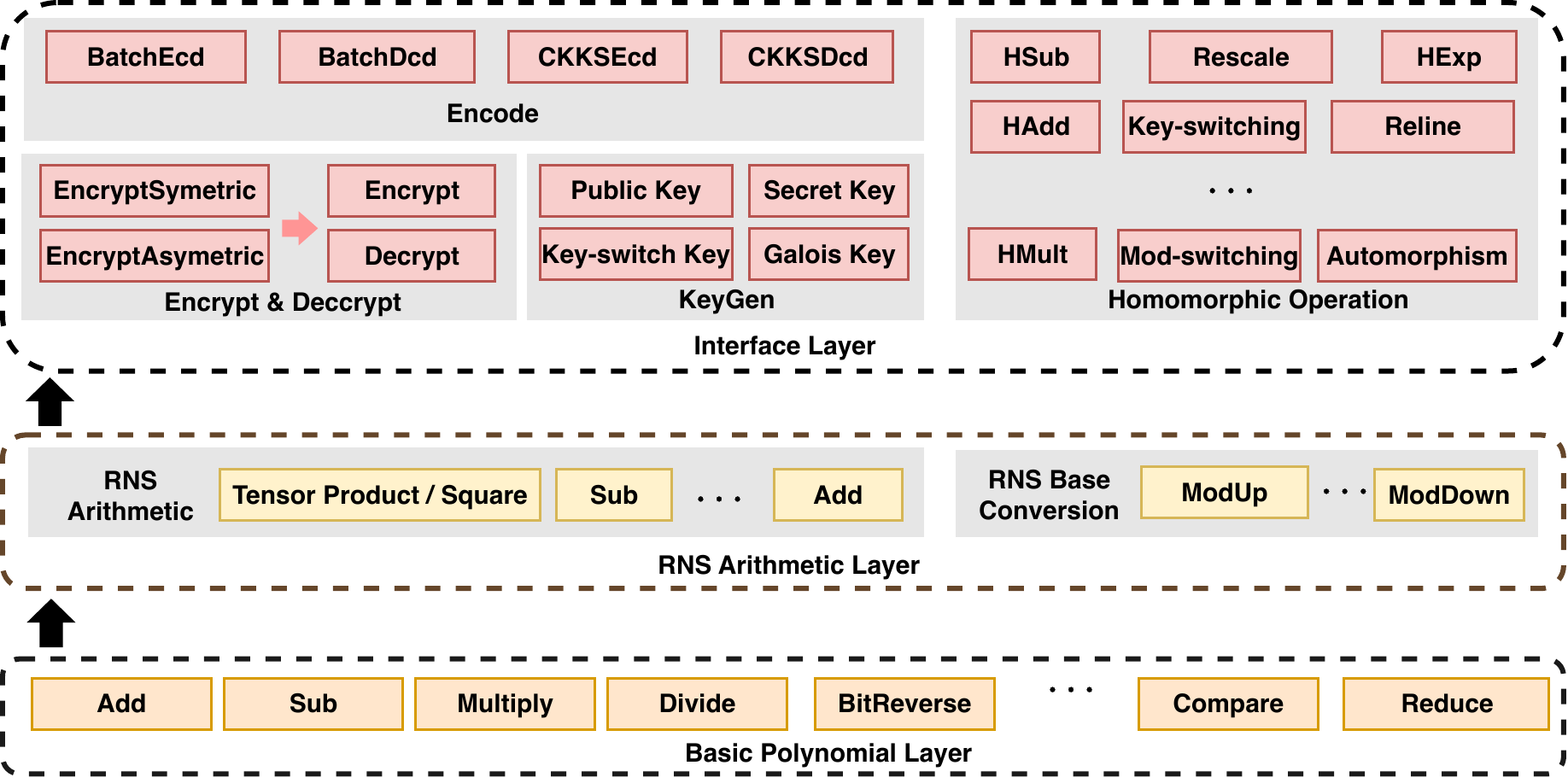}
\caption{Overview of Hardware Framework Structure.}
\label{fig:arch_lib}
\end{figure}

We show the structure of the implemented framework in Fig.~\ref{fig:arch_lib}. Our computational framework is meticulously architected into a hierarchical tri-layered design, encompassing a foundational Basic Polynomial Layer that includes low-level modular operations for 64-bit and 128-bit integers, a pseudo-random number generator, and sophisticated polynomial arithmetic. At the intermediate stratum, the RNS Arithmetic Layer is dedicated to polynomial operations under RNS or double-CRT representations, facilitating efficient computation through modules for sampling and advanced arithmetic. Culminating in the Interface Layer, it offers a high-level, cohesive implementation of complex encryption schemes, such as CKKS. The framework's integer operations are deftly implemented in CUDA PTX assembly to optimize instruction count and register utilization. This optimization is the bedrock for accelerated polynomial arithmetic using NWT and FFT algorithms. Comprehensive in its capabilities, our framework supports symmetric and asymmetric encryption, homomorphic operations including addition, subtraction, and multiplication of plaintexts and ciphertexts, as well as in-place homomorphic exponentiation, negation, and rotation of ciphertexts, catering to the nuanced requirements of secure, homomorphic encryptions.

\subsubsection*{Hardware Acceleration on Homomorphic Multiplication}
Due to the most time-consuming part of homomorphic encryption is homomorphic multiplication, we introduce how we use the GPU to accelerate multiplication in ciphertext. Given a two ciphertexts $\texttt{ct} = (\bm{c}_{0}, \bm{c}_{1})$, $\texttt{ct}' =(\bm{c}_{0}', \bm{c}_{1}')$ in $R_{Q_\ell}^{2}$, we have
$$\mathsf{HMult}(\texttt{ct}, \texttt{ct}' )=  (\bm{d_0}, \bm{d_1}, \bm{d_2}) = (\tilde{\bm{c}}_{0}, \tilde{\bm{c}}_{1}, \tilde{\bm{c}}_{2})=[(\bm{c}_{0}\cdot\bm{c}_{0}', \bm{c}_{0}\cdot\bm{c}_{1}' + \bm{c}_{1}\cdot\bm{c}_{0}', \bm{c}_{1}\cdot\bm{c}_{1}')]_{Q_\ell}.$$ 

Since we apply the CKKS scheme in \workname, we keep the ciphertext under double-CRT representation, there is no need for NWT/INWT transformation before
and after the tensor product.  In the tensor product kernel, we account for all potential scenarios, which vary according to the size of the input ciphertexts. The preferred configuration is when each ciphertext comprises two ring elements, as larger sizes lead to increased computational and memory demands. Similar to For our implementation, we employ the Karatsuba algorithm~\cite{karatsuba1962multiplication,yang2023implementing}, to calculate intermediate results of homomorphic multiplication. In particular,  the classical approach to obtain $\bm{d_1} =  [\bm{c}_{0}\cdot\bm{c}_{1}']_{Q'} + [\bm{c}_{1}\cdot\bm{c}_{0}']_{Q'}$. In Karatsuba method, we first compute $ \bm{d_1} = [(\bm{c}_{0} + \bm{c}_{1}) \cdot( \bm{c}_{0}' + \bm{c}_{1}')]_{Q'}$ and then compute $\bm{d_1}  = [\bm{d_1} - \bm{d_0} - \bm{d_2} ]_{Q'}$. The primary advantage of this approach is that it derives the 128-bit product result from two 64-bit CUDA PTX instructions. The Karatsuba method effectively reduces computational overhead by replacing certain 64-bit multiplication instructions with less costly addition and bitwise shift-right operations for each coefficient.

\subsection*{Related Works}

\subsubsection*{Privacy-Preserving Byzantine-Robust FL}

To enhance the robustness of FL under malicious client threats, several byzantine-robust FL methods have been proposed such as Krum~\cite{blanchard2017machine}, Multi-Krum~\cite{Blanchard2017MachineLW}, RFA~\cite{pillutla2022robust}, Trim-mean~\cite{Yin2018ByzantineRobustDL}, Median~\cite{Yin2018ByzantineRobustDL}.  For example, the Krum~\cite{blanchard2017machine} algorithm chooses one local update that minimizes the Euclidean distance to all other updates, and the Median method~\cite{Yin2018ByzantineRobustDL} calculates the median of all parameters to resist Byzantine attacks. 
However, these schemes have limitations on privacy since they analyze and calculate gradients in plaintext, leading to potential privacy leakage. To relieve such information leakage, privacy-preserving techniques such as differential privacy (DP), Secure multi-party computation (MPC), and homomorphic encryption (HE) are well adopted in byzantine-robust FL. In particular, Some recent work also investigates the problem of maintaining both DP and Byzantine resilience in federated learning \cite{guerraoui2021combining, Guerraoui2021DifferentialPA, zhu2022bridging}.  Specifically, DP-RSA~\cite{zhu2022bridging} bridges differential privacy and robust stochastic aggregation by sending the element-wise signs of the differences between their local and the global model. The following approaches~\cite{Guerraoui2021DifferentialPA,guerraoui2021combining} combines DP-SGD with existing aggregation rules on the noisy gradient. As for adopting MPC to Byzantine-Robust FL~\cite{nguyen2022flame,rathee2023elsa,gehlhar2023safefl}, He et al.~\cite{he2020secure} directly integrate secure computation based on additive secret sharing and a variation of the Krum aggregation protocol. In addition, So et al.~\cite{so2020byzantine} developed a similar scheme based on the Multi-Krum aggregation rule. However, they employed  verifiable Shamir's secret sharing and a robust gradient descent approach, enabled by secure computations over the secret shares of the local updates. 


\subsubsection*{HE-based Byzantine-Robust FL}
Several works have applied HE in privacy-preserving byzantine-robust FL~\cite{ma2022shieldfl,Dong2021FLODOD,Liu2021PrivacyEnhancedFL,rahulamathavan2023fhefl,nie2023bpfl}. In particular, ShieldFL~\cite{ma2022shieldfl} proposed a
two-trapdoor homomorphic encryption approach to encrypt local gradients and use the cosine similarity technique to resist poisonous gradients among encrypted gradients on the server side. 
FLOD~\cite{Dong2021FLODOD} introduces a Hamming distance-based aggregation method to resist Byzantine attack which is combined with MPC and HE-based protocols to construct efficient privacy-preserving aggregation. PEFL~\cite{Liu2021PrivacyEnhancedFL} adopts HE as
the underlying technology and provides the server with a channel to punish poisoners via the effective gradient data extraction of the logarithmic function to resist Byzantine attack. Unfortunately, the aforementioned methods merely consider the applying HE on byzantine-robust FL from the computation efficiency.  In FheFL~\cite{rahulamathavan2023fhefl}, each user employs a shared secret key to encrypt their model updates using the CKKS scheme and transmits these encrypted updates to the server and the server can aggregate the model updates from all users within the encrypted domain.
BPFL~\cite{nie2023bpfl} validates client models with zero-knowledge proofs and uses homomorphic encryption to protect client data privacy to achieve a unified Byzantine-robust FL system. Unfortunately, the aforementioned methods merely consider applying HE on byzantine-robust FL from the computation efficiency perspective.

\subsubsection*{Accelerations on CKKS}
Several works leverage hardware accelerators to improve the efficiency of FHE schemes such as GPU, FPGAs. In this work, we focus on using GPU to accelerate CKKS due to the wider accessibility and cost-effectiveness. 
The acceleration of polynomial arithmetic in CKKS typically involves two primary operations, the Chinese Remainder Theorem (CRT) and Number-Theoretic Transform (NTT). 
Privf~\cite{al2020privft} provide the first GPU implementation of a
CRT variant of the CKKS scheme and demonstrate how to perform both inference and training on encrypted inputs for text classification. Jung et al.~\cite{jung2021accelerating} demystified the key operations of CKKS and effectively exploits massive thread-level parallelism to GPUs by applying transposing matrices and pinning data to threads. Soon, Jung et al. in another work~\cite{Jung2021Over1F} employed two kernels to accelerate NTT, extending the acceleration to the CKKS scheme with significant parameters. In addition, they identified that the main memory bandwidth bottleneck is a key challenge in accelerating CKKS operations using GPUs. 
CARM~\cite{Shen2023CARMCR} also employed two kernels to accelerate NTT and is the first optimized GPU implementation of FHE schemes designed specifically for Internet of Things (IoT) scenarios. In addition, HE-booster~\cite{wang2023he} presents a systematic acceleration. They propose the local synchronization to allow threads to be executed as early as possible to accelerate NTT. They also propose a scalable partition strategy for multiple GPU, exploiting the tremendous data level parallelism. 






\subsection*{Descriptions of Aggregation Rules}
In non-adversarial environments, a straightforward approach to aggregating local model parameters into global model parameters is simply to take the average~\cite{dean2012large,mcmahan2017communication}. However, this simplistic aggregation rule is prone to manipulation in adversarial environments. Specifically, an attacker could arbitrarily manipulate the global model parameters by compromising just a single worker device. Recognizing this vulnerability, the machine learning community has recently developed a variety of aggregation rules designed to maintain robustness in the face of potential Byzantine failures among worker devices. In the following, we will review three of these aggregation rules.

\noindent \textbf{Krum~\cite{blanchard2017machine}.} Krum aims to select a single model from the pool of $n$ local models, specifically favoring those that demonstrate high degrees of congruity with others in the set. The intuition is rooted in a protective strategy for the global model. Even if the chosen local model is derived from a compromised worker device, the potential negative impact on the overall system is mitigated due to the chosen model's resemblance to other local models. In particular, the global model update is based on a square distance score. For example, suppose that $n$ local models $W_1, W_2, \ldots, W_n$, and at most $c$ clients are compromised. The score can be computed as follows.
$$
  s_i = \sum_{j \in \phi} \lVert W_j - W_i \rVert^2_2  
$$
where, $\lVert \cdot \rVert^2_2$ denotes the Euclidean distance and $\phi$ is the set of $n-c-2$ updated local models that have the smallest Euclidean distance to $W_i$.   Note that Krum has theoretical guarantees for the convergence of certain objective functions when $c < \frac{n-2}{2}$.

\noindent \textbf{Multi-Krum~\cite{Blanchard2017MachineLW}.} Multi-Krum extends the Krum algorithm to select multiple local model updates rather than one. In particular, the Multi-Krum relies on a similar scoring mechanism to select the most representative models. It begins with an empty set, $\psi$, and applies the Krum algorithm iteratively to the remaining local model updates, adding the one with the smallest Krum score to $\psi$. This process continues until $\psi$ contains $l$ updates, satisfying the condition $n - l > 2c + 2$, where $n$ is the number of clients, and $c$ is the number of compromised clients. The final global model update is obtained by averaging the updates in $S$.

\noindent \textbf{Median~\cite{blanchard2017machine}.} Median is a coordinate-wise method. The FL server picks the medians of each coordinate from all the weight updates to build the global weights. Given a set of weight updates $W_t$, at a communication round t, the FL server uses the coordinate-wise Median to do $Cod^{j}(W_t)$ with aggregation, where $Cod^{j}(W_t)$ is a vector with its $j$-th coordinate of $W_t$. However, in this paper, we modified the Median method by distance median due to the limitation of FHE. In particular, following the Krum, we use the median client as our aggregated model.

\begin{algorithm}[!ht]
    \caption{The Overall Workflow of \workname}
    \begin{algorithmic}[1]
    \REQUIRE  Weight $W_i$ from $i$-th client, $R$ $\in$ $\{$Krum, Multi-Krum, Median$\}$, compromised clients $c$, loss function $L$, labeled data $X$, $Y$, learning rate $\eta$. 
    \ENSURE An aggregated model $\mathbf{W}^*$ under FHE protection.
    \STATE \textbf{Key Generation Center (KGC) Executes:}
    \STATE \quad  Initialize the encryption parameters, weight $\mathbf{W_0}$.
    \STATE \quad  Generates the $\pk$, $\sk$,  and sends the $\pk$ to each client.  
    \STATE \quad \textbf{For} each round $t = 1,2, \cdots, N$ \textbf{do}
    \STATE \quad \quad $\llbracket W_{t} \rrbracket$ $\leftarrow$ \textbf{ClientUpate($W_t$)}
    \STATE \quad \quad $\llbracket SumDis \rrbracket$ $\leftarrow$ \textbf{Server Executes( $\llbracket W_{t} \rrbracket$)} 
    \STATE \quad \quad $SumDis$ $\leftarrow$  $\mathsf{Dec}_\sk$($\llbracket SumDis \rrbracket$) \hfill $\triangleright$ \textit{KGC decrypts distance and generate Mask.}
    \STATE \quad \quad $Mask$ $\leftarrow$ Masked-based Encrypted Sorting($SumDis$, $R$) 
    \STATE \quad \quad $\llbracket W_{t+1} \rrbracket$ $\leftarrow$ \textbf{ServerUpdate($Mask,R,c$)} 
    \STATE \quad \quad $ W_{t+1}$ $\leftarrow$  $\mathsf{Dec}_\sk$($\llbracket W_{t+1} \rrbracket$)  \hfill $\triangleright$ \textit{KGC decrypts the aggregated model and distributes it to all clients.}
    \STATE \quad $\mathbf{W}^*$ $\leftarrow$ $W_{N}$

    \STATE
    \STATE  \textbf{Server Executes($\llbracket W_{t}\rrbracket$}): \hfill $\triangleright$ \textit{Compute the distance  with Lazy Relinearization and Dynamic Hoisting.}
    \STATE \quad \textbf{For} $i = 1, 2, \cdots n$ \textbf{do}
    \STATE  \quad \quad  $ LEN  \leftarrow len (\llbracket W_{t}^{i}\rrbracket )/ N_{slot} $
    \STATE \quad \quad \textbf{For} $j = i, \cdots n$ \textbf{do}    
    \STATE \quad \quad \quad \textbf{For} $k = 1, \cdots LEN$ \textbf{do}
    \STATE \quad \quad \quad \quad  $T_{sub}$ $\leftarrow$ $\llbracket W_{t}^{i}\rrbracket$ $[k*N_{Slot}]$ $\ominus$ $\llbracket W_{t}^{j}\rrbracket$ $[k*N_{Slot}]$
    \STATE  \quad \quad \quad \quad $T_{mul}$ $\leftarrow$ $T{sub}$ $\otimes$ $T{sub}$
    \STATE  \quad \quad \quad \quad $Sum_{Dis}$ $\leftarrow$  $Sum_{Dis}$ $\oplus$ $T_{mul}$
    \STATE \quad \quad \quad $\llbracket Dis[i][j]\rrbracket$  $\leftarrow$  Relinearization($Sum_{Dis}$)
    \STATE \quad \quad \quad $\llbracket SumDis[i]\rrbracket$  $\leftarrow$ $\llbracket SumDis[i]\rrbracket$ $\oplus$  $\llbracket Dis[i][j]\rrbracket$

    \STATE
    \STATE  \textbf{ServerUpdate($Mask, R, c$)}: \hfill $\triangleright$ \textit{Aggregates the models using Mask and sends the model to the KGC.}

    \STATE \quad   \textbf{For} $j = 0, 1, \cdots, c$ \textbf{do} 
    \STATE \quad \quad  \textbf{For} $i = 0, 1, \cdots, n-1$ \textbf{do}
    \STATE \quad \quad  \quad  $\text{T} \leftarrow  W_{t}^{i}$ $\otimes$ $Mask[0][i]$
    
    \STATE \quad \quad \quad  $Chosen\_W\_list[j]$ $\leftarrow$  $Chosen\_W\_list[j]$ $\oplus$  \text{T}
   
    \STATE \quad  $Chosen\_W$ $\leftarrow$  $\sum$ $Chosen\_W\_list$ $\otimes$ $(c/n)$   \hfill $\triangleright$ $\textit{$R$ is Multi-Krum}$

    \STATE \quad  \textbf{For} $i = 0, 1, \cdots, n-1$ \textbf{do}
    \STATE \quad \quad  $T$ $\leftarrow $ $W_{t}^{i}$ $\otimes$ $Mask[0][i]$ 
    \STATE \quad \quad $Chosen\_W$ $\leftarrow$ $Chosen\_W$ $\oplus$ $T$  \hfill $\triangleright$ \textit{$R$ is Krum or Median}
    \STATE \quad  $\llbracket W_{t+1} \rrbracket$ $\leftarrow$ $Chosen\_W$
    \STATE
    \STATE   \textbf{ClientUpate($\mathbf{W}_{t}$)}:  \hfill $\triangleright$ \textit{Encrypt and Transmit Models}.  
    \STATE \quad \textbf{For} each client $k = 1,2, \cdots, n$ \textbf{do} (in parallel)
    \STATE \quad \quad $W_{t}^{k}$ $\leftarrow$ $W_{t}^{k}$ - $\eta\nabla L$($W_{t}^{k}$, $X$, $Y$)
    \STATE \quad \quad $\llbracket W_{t}^{k} \rrbracket$ $\leftarrow$       $\mathsf{Enc}_\pk$($W_{t}^{k}$)
    \STATE \quad $\llbracket W_{t} \rrbracket$ = $\cup_k$ $\llbracket W_{t}^{k} \rrbracket$

    \end{algorithmic}
    \label{alg:lancelot}
\end{algorithm}

\begin{algorithm}[!ht]
    \caption{An overview of Masked-based Encrypted Sorting}
    \begin{algorithmic}[1]
    \REQUIRE  $SumDis,R$ 
    \ENSURE An encrypted $Mask$

    \STATE  $SumDisIndex \leftarrow$ ArgSort($SumDis$)  \hfill $\triangleright$ \textit{Obtain the index according to the sorting result}
    \STATE  $SelectedId \leftarrow$  $SumDisIndex[0]$  \hfill $\triangleright$ \textit{$R$ is Krum}

    \STATE   $SelectedId \leftarrow$ $n \% 2 == 0$? $SumDisIndex[n/2]$ : $SumDisIndex[(n+1)/2]$   \hfill $\triangleright$ 
     \textit{$R$ is Median}
     
    \STATE   Initialize an empty set $\psi$ \hfill $\triangleright$ \textit{$R$ is Multi-Krum}
    \STATE  \textbf{While}{$|\psi| < l$ and $n - |\psi| > 2c + 2$}:
    \STATE  \quad $SelectedId \leftarrow SumDisIndex[0]$  \hfill $\triangleright$ \textit{Select the first index from the sorted distance index array.}
    \STATE  \quad Add $SelectedId$ to $\psi$
    \STATE  \quad Remove $SelectedId$ from $SumDisIndex$
    \STATE    $P_\pi$ $ \leftarrow$   $SelectedId \leftarrow$   $\psi$ \hfill $\triangleright$  \textit{Return the set of selected indices}

    \STATE  $Mask$ $\leftarrow$ $\Phi(P_\pi) = \{ \mathsf{Enc_{pk}}(\mathbf{e}_{j})| j \in n \}$.
    \end{algorithmic}
    \label{alg:mask-based-sorting}
\end{algorithm}

\end{document}